\documentclass{article}

\usepackage{arxiv}
\usepackage{amsmath}
\usepackage[utf8]{inputenc} % allow utf-8 input
\usepackage[T1]{fontenc}    % use 8-bit T1 fonts
\usepackage{hyperref}       % hyperlinks
\usepackage{url}            % simple URL typesetting
\usepackage{booktabs}       % professional-quality tables
\usepackage{amsfonts}       % blackboard math symbols
\usepackage{nicefrac}       % compact symbols for 1/2, etc.
\usepackage{microtype}      % microtypography
\usepackage{lipsum}		% Can be removed after putting your text content
\usepackage{graphicx}
\usepackage{natbib}
\usepackage{doi}

\title{Comparative Study of State-based Neural Networks for Virtual Analog Audio Effects Modeling}

\date{} 					% Or removing it

\author{ {\hspace{1mm}Riccardo Simionato} \\
	Department of Musicology\\
	University of Oslo\\
	Oslo, Norway \\
	\texttt{riccardo.simionato@imv.uio.no} \\
	\And
	{\hspace{1mm}Stefano Fasciani} \\
	Department of Musicology\\
	University of Oslo\\
	Oslo, Norway \\
	\texttt{stefano.fasciani@imv.uio.no} \\
}

\renewcommand{\shorttitle}{ }
\newcommand{\headeright}{ }
\newcommand{\undertitle}{ } % Empty definition

\hypersetup{
pdftitle={A template for the arxiv style},
pdfsubject={q-bio.NC, q-bio.QM},
pdfauthor={David S.~Hippocampus, Elias D.~Striatum},
pdfkeywords={First keyword, Second keyword, More},
}

\begin{document}
\maketitle

\begin{abstract}
Artificial neural networks are a promising technique for virtual analog modeling, having shown particular success in emulating distortion circuits. Despite their potential, enhancements are needed to enable effect parameters to influence the network's response and to achieve a low-latency output. While hybrid solutions, which incorporate both analytical and black-box techniques, offer certain advantages, black-box approaches, such as neural networks, can be preferable in contexts where rapid deployment, simplicity, or adaptability are required, and where understanding the internal mechanisms of the system is less critical. In this article, we explore the application of recent machine learning advancements for virtual analog modeling. In particular, we compare State-Space models and Linear Recurrent Units against the more common Long Short-Term Memory networks, with a variety of audio effects. We evaluate the performance and limitations of these models using multiple metrics, providing insights for future research and development. Our metrics aim to assess the models' ability to accurately replicate the signal's energy and frequency contents, with a particular focus on transients. The Feature-wise Linear Modulation method is employed to incorporate effect parameters that influence the network's response, enabling dynamic adaptability based on specified conditions. Experimental results suggest that Long Short-Term Memory networks offer an advantage in emulating distortions and equalizers, although performance differences are sometimes subtle yet statistically significant. On the other hand, encoder-decoder configurations of Long Short-Term Memory networks and State-Space models excel in modeling saturation and compression, effectively managing the dynamic aspects inherent in these effects. However, no models effectively emulate the low-pass filter, and Linear Recurrent Units show inconsistent performance across various audio effects.
\end{abstract}

\section{Introduction}\label{sec1}

The emulation of analog musical devices has become pervasive in digital products due to their distinct sound, appealing to producers and musicians. Analog circuits use components such as operational amplifiers and diodes whose non-linear behaviors contribute to creating a unique sound. Replicating these devices, called Virtual Analog (VA) modeling, is a crucial area in digital audio signal processing.

There are two possible approaches for emulating analog audio effects: physics-based and data-driven. Physics-based approaches discretize existing continuous-time models. Data-driven approaches identify a discrete model directly from discrete measurement data. Machine learning (ML), as a data-driven method, has become widely used in audio modeling and is often integrated into more traditional digital signal-processing frameworks. Hybrid solutions, incorporating knowledge of the modeling tasks, also exist and can offer advantages, but data-driven approaches can still be appealing in some contexts. While the former may not always be accurate enough, the latter may not be feasible for certain tasks. 
    
Data-driven models, also called Black-box models, have been utilized for different audio modeling tasks ~\cite{ramirez2020deep}, but these usually involve large networks that are not suitable for low-latency and real-time applications. In addition, while machine learning modeling of nearly all types of analog effects has been investigated, they often do not provide a satisfactory response when continuously manipulating conditioning parameters, which is needed to offer users runtime control over the effect's parameters.

Neural networks incorporating internal states and utilizing recurrent formulations have emerged as the most accurate option for modeling audio effects among various types of artificial neural networks, with Recurrent Neural Networks (RNNs) being the most recognized. A similar formulation exists in State-Space Models (SSMs). The internal states in these architectures serve as memory, rendering them particularly effective in handling scenarios with transients and time dependencies, especially when processing one or a few audio samples at a time to achieve low latency. Consequently, in this study, we include Convolutional Neural Networks (CNNs)~\cite{lecun1998gradient}---which are inherently stateless---only when integrated with RNNs~\cite{simionato2022deep}. 

This article presents a comprehensive comparative study that systematically evaluates state-based neural network architectures in modeling various types of audio effects, each posing unique modeling challenges, and in the context of achieving low-latency, computationally sustainable, and parametric solutions. Experimental results highlight the strengths and limitations of different state-based networks in their ability to learn diverse audio effects. Specifically, we explore the learning capabilities of these architectures, investigating the similarities and differences between them. By analyzing RNNs and SSMs, which use distinct recurrence methods to compute internal states, we provide insights into their variant modeling capabilities across different contexts. While architecture-specific optimizations, such as hyperparameter tuning or layer size adjustments, were not pursued, our study prioritized limiting computational complexity and minimizing input-output latency. We focused on developing models that are suitable for implementation on consumer-grade devices for live audio applications.

We designed five architectures based on the popular Long Short-Term Memory (LSTM) ~\cite{hochreiter1997long}, the SSMs ~\cite{gu2021efficiently, gu2023mamba}, and the Linear Recurrent Unit (LRU) ~\cite{orvieto2023resurrecting}. All models consider conditioning on control parameters using the Feature-wise Linear Modulation (FiLM) method ~\cite{perez2018film} and employ the Gated Linear Unit (GLU) ~\cite{dauphin2017language}. The architectures are engineered to operate with a computational load of less than $100$ MFLOPS (Million Floating Point Operations Per Second) when processing audio at $48$ kHz. This constraint facilitates the simultaneous operation of multiple software audio effects on consumer-grade personal computers and supports independent execution on bare-metal embedded systems. Additionally, the architectures are designed to exhibit minimal inherent input-output latency for audio samples. %, which we have limited to a maximum of $64$ samples ($1.33$~ms). 
Specifically, we refer to the intrinsic latency given by the architecture constraints, neglecting the latency given by the external Digital Signal Processor (DSP) processing. This latency is due to the lag between the time index of the input and output samples.

Section ~\ref{sec:VA} reviews the recent advancements in modeling analog devices, with a specific focus on the benefits and drawbacks of the recent deep learning methods against other data-driven methods. Section ~\ref{sec:methods} details the methodology, architectures, losses, and datasets used in our experiments. Section ~\ref{sec:results} presents the obtained results and the consequent discussion on the limit and strength of the current methodology. 

\section{Background}\label{sec:VA}

Physics-based modeling for analog audio effects requires comprehensive knowledge of the circuit schematics and components. These circuits need to be mathematically modeled and solved. A numerical scheme is formulated, and the discrete equations are solved iteratively to generate the digital audio output.

Data-driven modeling is typically used when the internal circuitry of an audio effect is unknown, treating the model as a mapping of input-output data to approximate the system's response. This process involves three main components: the dataset, modeling techniques, and validation methods. The model's quality is heavily dependent on the selection of input signals, control parameter configurations, and data collection strategies. However, because data-driven approaches often rely on static measurements, accurately capturing dynamic behavior presents a notable challenge.
    
Common examples of data-driven approaches include the block-oriented Hammerstein model, the Wiener model, and their combinations ~\cite{schoukens2017identification}. Recent examples of their application focus on guitar distortion pedal ~\cite{eichas2015block} ~\cite{koper2020taming} ~\cite{darabundit2019digital}, amplifier ~\cite{schuck2016audio} ~\cite{eichas2017block}, distortion circuit ~\cite{eichas2016black}.
    
More recent data-driven techniques leverage artificial neural networks (ANNs), comprising interconnected nonlinear units known as neurons. Each neuron propagates signals through its connections, processes these signals, and has the ability to transmit the processed signals to other connected neurons. Training ANNs involves optimizing the network's parameters to minimize the error between predicted and actual target values in a dataset. Gradient-based techniques, like backpropagation, are commonly used to optimize the network's weights.

\subsection{Neural Networks for Virtual Analog Modeling}

%DL
In the last two decades, ANNs have been extensively used for modeling analog audio devices. In particular, traditional ML techniques have been investigated to create models that emulate audio effects processing raw audio samples, such as traditional DSP techniques.
    
%FC
In an early attempt using ML for audio effect modeling ~\cite{mendoza2005emulating}, a multilayer feedforward network was utilized to learn a high-pass filter and the Ibanez Tube-Screamer distortion effect. The high-pass filter experiments led to unsatisfactory results as the model did not learn any aspect of the filtering process, while with the distortion effect, the network demonstrated the ability to replicate the signal alteration typical of such audio effects accurately. 

%CNN
Other studies utilized CNNs to model the preamplifier circuit of the Fender Bassman 56F-A vacuum-tube amplifier ~\cite{damskagg2019deep}. The proposed network features stacked dilated causal convolution and fully connected layers, trained on a synthetic dataset with varying gain values for conditioning the network through a gated activation technique ~\cite{oord2016wavenet}. Results indicate that CNNs outperform fully connected networks in emulating the device and they can also adapt to changes in the effect's control parameters. The same architecture was later applied to model the Ibanez Tube Screamer, Boss DS-1, and Electro-Harmonix Big Muff Pi distortion pedals ~\cite{damskagg2019real}, exploring various configurations without considering effects parameters for conditioning the network.
  
%TCN CNN
Similarly to fully connected networks, CNNs' ability to track the input's past information depends on their receptive field, which requires more layers when the fields are longer. Temporal Convolutional Networks (TCNs) use dilated convolutions with increasing dilation factors ($2^2$ and $2^{10}$), enhancing the receptive field and long dependency handling without increasing layer count. TCNs have been applied to optical dynamic range compressors, such as the Teletronix LA-2A Leveling Amplifier ~\cite{steinmetz2021efficient}, where the utilization of a $300$ ms receptive field yielded the best performance. TCNs provide accurate and computationally efficient implementations, but network architectures with convolutional layers present intrinsic input-output latency equal to the receptive field length, which can hinder live audio applications. TCNs have also been used to design Open-Amp ~\cite{Wright25Open}, a model emulating many guitar effects. The model is conditioned on a learnable look-up table of embeddings and learns one embedding vector per effect. In this case, the length of the receptive field is not reported.

%hybrid
A model combining various deep learning architectures was explored in ~\cite{ramirez2019modeling}. The model includes an adaptive front-end, a latent space, and a synthesis back-end. The front-end uses a convolutional encoder to perform time-domain convolutions on raw audio, mapping it into a latent space. The latent-space transforms the input audio representation, which is then processed by the synthesis back-end, consisting of an unpooling layer, a deep neural network with smooth adaptive activation functions (SAAF) ~\cite{hou2017convnets}. This approach has been applied to model distortion, overdrive, and equalizer effects. 

In ~\cite{ramirez2020deep}, the previously described architecture was further investigated and compared with variations and the CNN-based architecture from ~\cite{damskagg2019deep}. Variations included incorporating Bi-LSTMs or CNNs into the latent space. The Bi-LSTM variation achieved the highest score in listening tests. Another Bi-LSTM and SAAF variation was used for plate and spring reverberators ~\cite{ramirez2020modeling}. While these architectures effectively model various audio effects, they are large networks with high computational complexity, making efficient implementation challenging. The use of Bi-LSTMs also introduces non-causality, hindering real-time applications and interactivity with the effect's parameters. In ~\cite{comunita2023modelling}, LSTMs are integrated to enhance intermediate data representation in TCN models, tested on fuzz and compressor effects, and the proposed architecture outperforms the baselines of LSTM and CNN. As in ~\cite{steinmetz2021efficient}, the models present long receptive fields, in this case, between $45$ ms and $2500$ ms. 
    
Finally, RNNs have been used to model vacuum-tube amplifiers ~\cite{covert2013vacuum}. Although preliminary experiments showed limited accuracy, they established RNN feasibility for this modeling task. RNNs, particularly LSTMs, were also compared to CNNs in modeling the Ibanez Tube Screamer, Boss DS-1, and Electro-Harmonix Big Muff Pi distortion pedals ~\cite{wright2019real}. The architecture included one RNN layer and a linear output layer, with LSTM and Gated Recurrent Unit (GRU) compared by varying the number of units. The input of the network includes both the audio signal and the effect parameters. The results indicated that RNNs achieved similar accuracy to CNNs, but with lower computational costs and latency, relying only on the current input sample for output prediction. LSTM outperformed GRU, with error reduction observed as unit count increased.
 
In ~\cite{chowdhury2020comparison}, another RNN-based model for the Klon Centaur guitar pedal circuit is compared with a Wave Digital Filter (WDF) model. The machine learning approach presented a lower computational cost, though the emulated sound was very similar to the WDF model, with a slight damping of high frequencies. The architecture includes a GRU layer and a linear output layer with one unit, and a single control parameter is also fed as input to the network to condition the inference.
    
In ~\cite{mikkonen2023neural}, a tape-based delay effect is modeled using GRU networks, with the delay trajectory analyzed through impulse train signals. This trajectory demodulates the signal before model training or guides a differentiable delay line based on ~\cite{engel2020ddsp}. In ~\cite{wright2021neural}, the modeling of delay-based effects, such as flanger and phaser, is investigated using an extracted low-frequency oscillation (LFO) signal to aid LSTMs during training. This is provided as additional input to the network, and it is shown to improve the learning of these effects. A the same time, it allows for post-training control of the model; however, the experiments were focused on fixed effect parameters, without changes in the LFO depth and width over time.

An encoder-decoder (ED) network was utilized to model the TubeTech CL 1B compressor ~\cite{simionato2022deep}, featuring an encoder LSTM that processes past input samples and control parameters, and a decoder LSTM handling current and more recent past input samples. The encoder's internal states summarize the input, aiding the decoder's output inference. This model outperformed standard LSTMs and fully connected networks in accuracy, initially modeling only threshold and ratio control parameters, while the attack and release settings were fixed. In ~\cite{simionato2023fully}, all four compressor control parameters were modeled, and the encoder was replaced with a convolutional layer for comparison with TCNs. While ED models offer increased responsiveness and interactivity, they also incur higher computational costs.

%Gray
Other studies explored hybrid modeling approaches. In ~\cite{parker2019modelling}, fully connected neural networks utilize circuit measurements, embedding the network within a discrete-time state-space system for diode-based guitar distortion circuits and the Korg MS-20 lowpass filter. This method relies on circuit measurements, which may not always be accessible. Similarly, ~\cite{peussa2021exposure} proposes a circuit state-matching mechanism for GRU networks, training a variant of the GRU layer to match circuit states using a teaching-forcing method.

The Differentiable Digital Signal Processing (DDSP) framework for modeling an RC lowpass filter is employed in ~\cite{kuznetsov2020differentiable}, later also used for the parametric tone section of a guitar amplifier and a nonlinear overdrive circuit in ~\cite{esqueda2021differentiable}. This method learns circuit parameters from real device measurements using a Wiener-Hammerstein model with two trainable Infinite Impulse Response (IIR) filters and a feedforward neural network. Similarly, ~\cite{nercessian2021lightweight} trains a cascade of biquad filters via the Fourier transform for the BOSS MT-2 distortion pedal, incorporating four tunable parameters. The DDSP framework was also successfully applied to a diode clipper circuit in ~\cite{chowdhury2022emulating}, demonstrating higher accuracy with comparable computation times to traditional models.

In ~\cite{wilczek2022virtual}, the authors learn ordinary differential equations governing diode clippers using fully connected layers and RNNs, aligned with the discrete-time state-space approach from ~\cite{parker2019modelling}. Investigations into WDF formulations are found in ~\cite{darabundit2022neural}, simulating tube amplifier circuits with multiport nonlinearities and fully connected layers. ~\cite{miklanek2023neural} combines RNNs with Kirchhoff nodal analysis to model guitar amplifiers. A hybrid approach for the LA-2A leveling amplifier is presented in ~\cite{graycomp}, modeled using a traditional digital compressor structure alongside fully connected RNNs for parameter prediction. Lastly, ~\cite{huhtala2024klann} proposed a Koopman-Linearised Audio Neural Network structure, which comprises two gated linear unit multilayer perceptrons and differentiable biquad filters.

\section{Methods}\label{sec:methods}

This study aims to identify which type of recurrent layer or architectural configuration accurately models different types of audio effects, if any. These insights can be used to understand the strengths and weaknesses of these models and determine which can better suit each audio effect. The state-based architectures offer advantages due to their ability to capture time dependencies in the data by utilizing internal states rather than solely relying on the input received at each iteration. In addition, we restrict the experimentation to relatively small networks that are fed by a limited number of input samples. This approach helps reduce computational complexity and minimizes input-output latency, enabling implementation on real-time consumer-grade digital audio systems.  

The designed models are applied to various audio effects and evaluated using a range of metrics. It is important to note that, given their limited number of trainable parameters, these models are not expected to achieve perfect emulation but rather serve as a solid foundation for further development.

In the rest of this section, we first detail the five architectures used in this comparative study; then, we present the metrics we used for evaluating the trained models and the specific audio effects from which we collected the datasets. Finally, explain the process we used to train and test the models.

\subsection{Architectures}\label{sec:nn}
\begin{figure}[h]%
\centering
\includegraphics[width=1\textwidth]{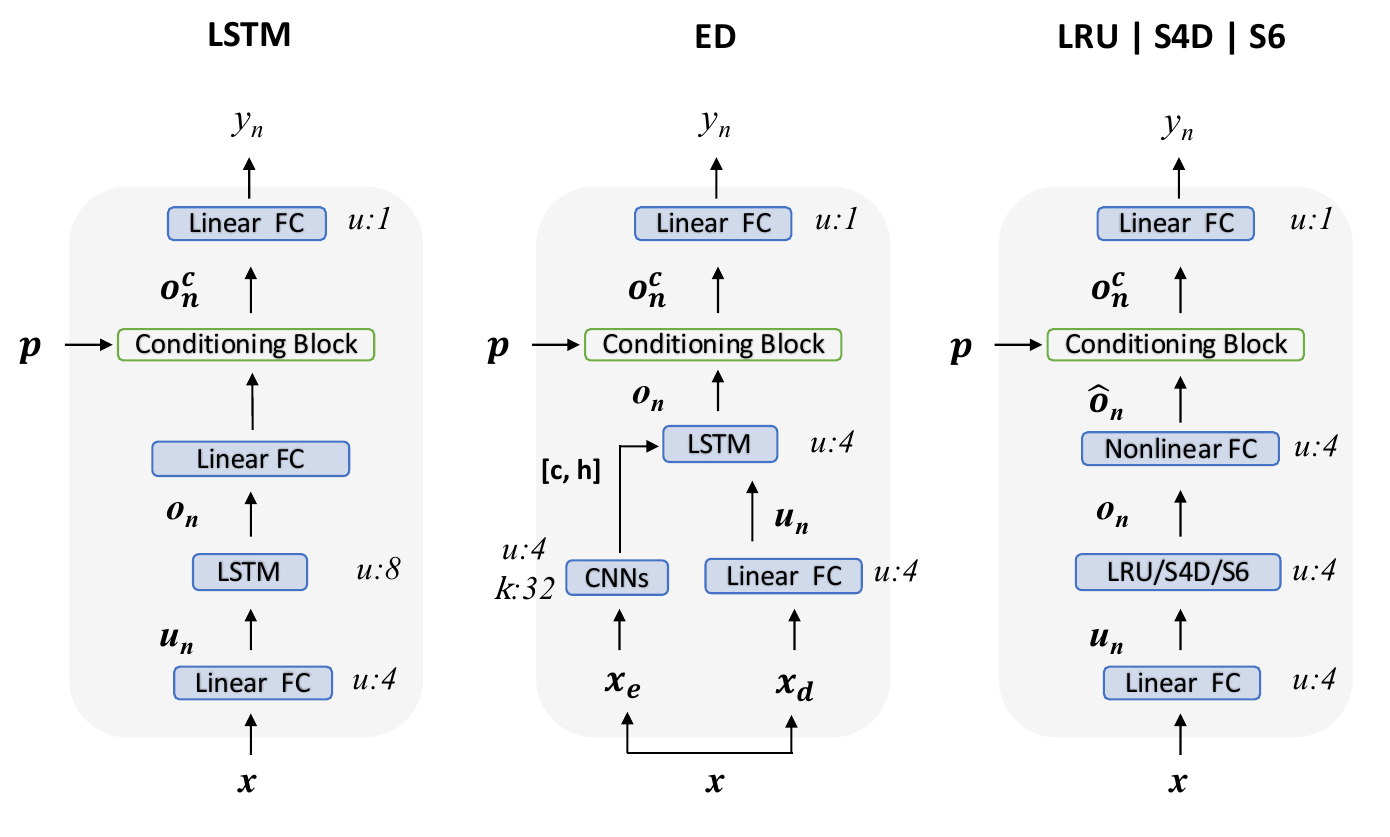}
\caption{The five compared architectures: \textbf{LSTM}, \textbf{ED}, \textbf{LRU}, \textbf{S4D}, and \textbf{S6}. All architectures present the same structure: a fully connected layer, a recurrent layer, and a conditioning layer. For the case of \textbf{ED}, the input vector is equally split into two vectors, $\boldsymbol{x_e}$ and $\boldsymbol{x_d}$. The vector $\boldsymbol{p}$ denotes the effect parameters utilized to condition the networks.
}\label{fig:arch}
\end{figure}

The following architectures, based on different types of recurrent layers, are compared in this study: \textbf{LSTM}, \textbf{ED}, \textbf{LRU}, \textbf{S4D}, and \textbf{S6}. The five architectures are illustrated in Figure~\ref{fig:arch}. \textbf{ED} is based on the concept of internal state sharing ~\cite{simionato2022deep}. These architectures follow the same design, consisting of a linear, fully connected layer, a recurrent layer, and a conditioning block.

The number of input samples is identical for all architectures and set to $64$. This choice represents a trade-off: on the one hand, there is the amount of past input information used to generate the current output, where more is typically better; on the other hand, there is the resulting input-output latency, which increases with the number of input samples. Our goal is to keep this latency below $1.33$~ms (i.e., $64$ samples at $48$ kHz). The input of all architectures is an array, including the $64$ most recent input sample, used to generate the single current output sample. This implies that a new input sample enters the buffer with each inference cycle while the oldest is discarded. Therefore, the set of the $64$ most recent input samples constitutes the network input, which, after initial compression by the FC layer, is used to update the states in the recurrent layers. This approach aids the network in making predictions while maintaining a stateful design using truncated backpropagation through time. Predicting one sample at a time helps minimize the audible artifacts commonly associated with machine-learning models of audio effects, which arise from slight amplitude mismatches at the boundaries of consecutive output segments. Consequently, the model must perform predictions at an audio rate, one for each output audio sample.
The models map a vector of $64$ samples into one scalar output sample, giving the internal states $\boldsymbol{h}$:
\begin{equation}
        y_n = g (\boldsymbol{x_n}) = g (x_n, x_{n-1}, ..., x_{n-62}, x_{n-63}; \boldsymbol{h})  \label{eq:general}
\end{equation}
where $n$ is the discrete time index, and $g$ is the overall function describing the models. 
Before being fed into the recurrent layer (LSTM, LRU, S4D, or S6), the input samples array $\boldsymbol{x_n}$ is processed by a fully connected layer that performs a linear projection:
\begin{equation}
    \boldsymbol{u_n} = \boldsymbol{W_l} \boldsymbol{x_n} + \boldsymbol{b_l} \label{eq:proj}
\end{equation}
where $\boldsymbol{u_n}$ is the linearly projected vector having a dimensionality equal to the number of hidden units of the fully connected layer. $\boldsymbol{W_l}$ and is the weights' matrices, and $\boldsymbol{b_l}$ the vector bias terms.
Following the linear projections, the models are characterized by various sets of equations, contingent upon the specific type of layer employed. 

\paragraph{LSTM layer} The vector $\boldsymbol{u_n}$ is processed using a LSTM as the recurrent layer. LSTMs were designed to improve the memory capacity of the vanilla RNN by incorporating internal mechanisms known as gates, which regulate the flow of information. The LSTM architecture is governed by the following equations:
\begin{align}
    \boldsymbol{f_n} &= \gamma(\boldsymbol{W}_f \boldsymbol{h_{n-1}} + \boldsymbol{U}_f \boldsymbol{u_{n}} + \boldsymbol{b}_f)  \nonumber\\
    \boldsymbol{i_n} &= \gamma(\boldsymbol{W}_i \boldsymbol{h_{n-1}} + \boldsymbol{U}_i \boldsymbol{u_{n}} + \boldsymbol{b}_i)  \nonumber\\
    \boldsymbol{o_n} &= \gamma(\boldsymbol{W}_o \boldsymbol{h_{n-1}} + \boldsymbol{U}_o \boldsymbol{u_{n}} + \boldsymbol{b}_o)  \nonumber\\
    \boldsymbol{c'_n} &= \phi(\boldsymbol{W}_c \boldsymbol{h_{n-1}} + \boldsymbol{U}_c \boldsymbol{u_{n}} + \boldsymbol{b}_c)  \nonumber\\
    \boldsymbol{c_n} &=  \boldsymbol{f_n} \circ \boldsymbol{c_{n-1}} + \boldsymbol{i_n} \circ \boldsymbol{c'_n}  \nonumber\\
    \boldsymbol{h_n} &= \boldsymbol{o_n} \circ \phi(\boldsymbol{c_n}) 
\end{align}
where $\boldsymbol{h_n}$ is the hidden state vector, $\boldsymbol{f}$ is the forget gate, $\boldsymbol{i}$ the input gate, and $\boldsymbol{o}$ the output gate, $\boldsymbol{c}$ is the cell state vector, and $\circ$ the element-wise product. The forget gate determines what information from previous steps should be kept or discarded. The input gate decides which information from the current step is relevant to add. The output gate determines the value of the next hidden state. Additionally, the cell state is updated by multiplying the forget vector and the input vector with new values that the neural network deems relevant. The forget vector and input vector are obtained using the sigmoid function, denoted by $\gamma$, which determines how much information should be kept. Values of $0$ in this vector indicate information to be discarded, while a value of $1$ indicates information to be kept. Furthermore, the $\phi$ represents the hyperbolic tangent function, used to keep the value of the cell state between $-1$ and $1$. Finally, $\boldsymbol{W}$ and $\boldsymbol{U}$ are the weights' matrices, and $\boldsymbol{b}$ the vector bias terms.

\begin{figure}[h]%
\centering
\includegraphics[width=0.4\textwidth]{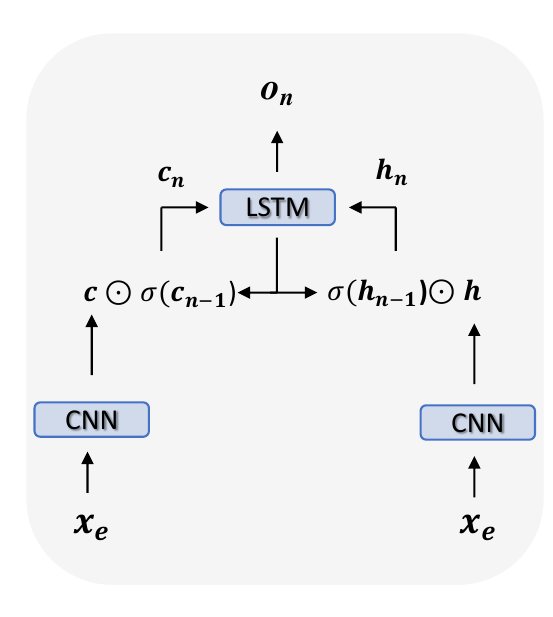}
\caption{The states sharing process in \textbf{ED}: the encoder input $\boldsymbol{x_e}$ is passed to two convolutional layers, which produce the states $[\boldsymbol{h}, \boldsymbol{c}]$. These states are combined with $[\boldsymbol{h_{n-1}}, \boldsymbol{c_{n-1}}]$, which are the LSTM states of the previous iteration. The sigmoid function is applied to $[\boldsymbol{h_{n-1}}, \boldsymbol{c_{n-1}}]$ and the results are element-wise multiplied to $[\boldsymbol{h}, \boldsymbol{c}]$. The outcome will be the new states for the LSTM layer $[\boldsymbol{h_n}, \boldsymbol{c_n}]$.}\label{fig:sharing}
\end{figure}

\paragraph{ED layer}
In this case, the input of the network $\boldsymbol{x}$ is split into two equal-size parts, $\boldsymbol{x_d}$ containing $[x_n, ..., x_{n-31}]$ and $\boldsymbol{x_e}$ containing containing $[x_{n-32}, ..., x_{n-63}]$. The first part feeds two separate convolutional layers, which compute the states, $[\boldsymbol{h}, \boldsymbol{c}]$. These states are combined with the internal states of the LSTM layer computed during the previous iteration. A sigmoid function is applied to the LSTM internal states and then element-wise multiplied by $[\boldsymbol{h}, \boldsymbol{c}]$. This operation determines how much information should be taken from the computed $[\boldsymbol{h}, \boldsymbol{c}]$. The result of this operation represents the new LSTM's internal states. This process is detailed in Figure~\ref{fig:sharing}.

\paragraph{LRU layer}
Compared to a vanilla RNN, the LRU uses a linear activation function to compute the hidden state instead of a nonlinear function. This modification allows the memory capability to improve significantly. The LRU equations are the following:
\begin{align}
    \boldsymbol{h_n} &= \boldsymbol{W_h} \boldsymbol{h_{n-1}} + \boldsymbol{U_h} \boldsymbol{u_n} + \boldsymbol{b_h}  \nonumber\\
    \boldsymbol{o_n} &= \boldsymbol{W_o} \boldsymbol{h_n} + \boldsymbol{b_o} 
\end{align}
where $\boldsymbol{h_n}$ is the hidden state at time $n$ and acts as 'memory' of the network, $\boldsymbol{o_n}$ the output vector. LRU keeps the recurrent layer linear but moves the nonlinearity to a subsequent fully connected layer. Additionally, to further enhance stability, the LRU incorporates a complex-valued diagonal recurrent matrix initialization and an exponential parameterization. At the same time, a normalization scheme for the hidden states is used to boost the efficiency and accuracy~\cite{orvieto2023resurrecting}.

\paragraph{S4D layer}
The SSMs operate by considering a continuous-time representation of the state-space formulation that is discretized to obtain the following equations:
\begin{align}
    \boldsymbol{h_n} &= \boldsymbol{\bar{W}_A} \boldsymbol{h_{n-1}} + \boldsymbol{\bar{W}_B} \boldsymbol{u_n}  \nonumber\\
    \boldsymbol{o_n} &= \boldsymbol{\bar{W}_C} \boldsymbol{h_n} + \boldsymbol{\bar{W}_D} \boldsymbol{u_n}. \label{eq:ssm}
\end{align}
In our study, we employ the diagonal state-space model variant S4D ~\cite{gu2022parameterization}. The remarkable ability of S4D to capture long-range dependencies stems from its utilization of a specific state matrix $\boldsymbol{\bar{W_A}}$ known as the "HiPPO matrix"~\cite{gu2020hippo}. This matrix is designed to encode the entire history of past inputs within $\boldsymbol{h_n}$, effectively mapping $\boldsymbol{u_n}$ to a higher-dimensional space $\boldsymbol{h_n}$ that encapsulates the compressed representation of this historical input data. The matrix allows the model to be conceptualized as a convolutional model that decomposes an input signal onto an orthogonal system of smooth basis functions. In this way, the state $\boldsymbol{h_n}$ encodes the history of the inputs. In the diagonal form, the S4D model is defined by parameterizing its state matrix as a diagonal matrix, boosting the model's efficiency. By doing so, the basis kernels have closed-form formulas represented by normalized Legendre polynomials $L_n(t)$, resulting in the SSM decomposing the input $\boldsymbol{u_n}$ onto an infinitely long set of basis functions that are orthogonal with respect to an exponentially decaying measure. 

\paragraph{S6 layer}
The S6 layer~\cite{gu2023mamba} introduces input-dependent matrices. Specifically, $\boldsymbol{\bar{W_B}}$ and $\boldsymbol{\bar{W_C}}$ become dependent on the input and computed using a linear FC layer, while $\boldsymbol{\bar{W_A}}$ and $\boldsymbol{\bar{W_D}}$'s parameters are independent of the input and learned during the training process. 
\paragraph{}
After the recurrent layer---whether LSTM, ED, LRU, S4D or S6---a FC layer reduces the dimensionality to $4$, ensuring uniformity across all model variations before the conditioning block. For the \textbf{LRU}, \textbf{S4D}, and \textbf{S6}, this FC layer includes a hyperbolic tangent activation function, compensating for the absence of nonlinear activation functions within these layers. Conversely, the \textbf{LSTM} model uses a linear activation function in the FC layer, as detailed in the following equations:
\begin{align}
    \boldsymbol{\hat{o}_n} &= tanh( \boldsymbol{W_{nl}} \boldsymbol{o_n} + \boldsymbol{b_{nl}}) \text{    if     } \textbf{LRU / S4D / S6} \nonumber\\
    \boldsymbol{\hat{o}_n} &= ( \boldsymbol{W_{nl}} \boldsymbol{o_n} + \boldsymbol{b_{nl}}) \text{    if     } \textbf{LSTM}
    \label{eq:nl}
\end{align}

The conditioning block is then applied just before the output layer, which consists of a linear fully connected layer with one unit, which is described in the following subsection. Placing the conditioning block after the recurrent layers rather than before has been found beneficial ~\cite{simionato2024conditioning}. This suggests that it is more beneficial for the networks to use the information given by the control parameters to project the output of the recurrent layer and determine the extent to which this information influences the final output rather than influencing the inference of the recurrent layer based on the control parameters.

Finally, the output layer, which is a fully connected layer with one unit, computes $y_n$:
\begin{equation}
    y_n = (\boldsymbol{W_{out}} \boldsymbol{o^{c}_n} + b_{out}) \label{eq:out}
\end{equation}
where $\boldsymbol{o^{c}_n}$ is the output of the conditioning block, as detailed in Figure ~\ref{fig:cond}. Since the output layer has one unit, $\boldsymbol{W_{out}}$ is a vector of the same dimensionality as $\boldsymbol{o^{c}_n}$.

The number of units is chosen to have approximately $800$ trainable parameters among the architectures and effects, since an identical number is achievable due to major differences in the layers. As a result, the selected \textbf{LSTM} and \textbf{ED} present $8$ units in the recurrent layer and $4$ in fully connected ones. \textbf{LRU}, \textbf{S4D} and \textbf{S6} present $12$ units in the recurrent layer, and $6$ in fully connected ones. The number of trainable parameters also depends on the dimensionality of the conditioning input $\boldsymbol{p}$, equal to the number of variable parameters in the datasets.

\subsubsection{Conditioning Block}

\begin{figure}[h]%
\centering
\includegraphics[width=0.4\textwidth]{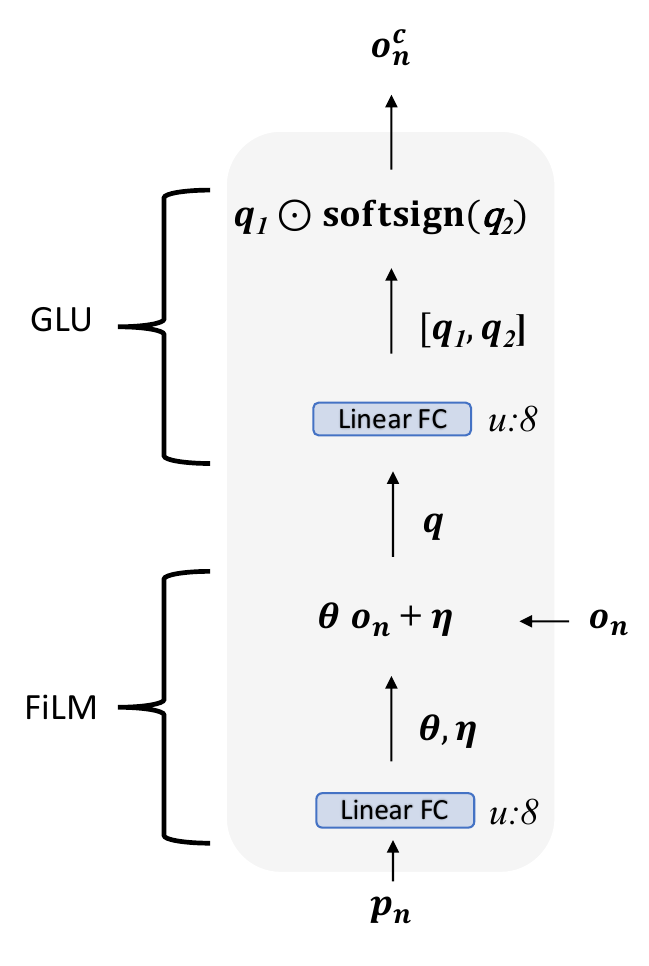}
\caption{Conditioning block based on the FiLM method and the GLU layer. The GLU layer features the softsign activation, $\boldsymbol{p}$ denotes the conditioning vector, and $\boldsymbol{o_n}$ represents the network's internal representation of the input.
}\label{fig:cond}
\end{figure}

The conditioning block consists of the FiLM method and the GLU layer, as in~\cite{simionato2024conditioning}, and is illustrated in Figure~\ref{fig:cond}. 

FiLM is a technique that applies an affine transformation to a vector based on conditioning values. The parameters vector $\boldsymbol{p}$, if applicable, is the input to a linear FC layer. The output of this layer is then split into two vectors of equal size: $\boldsymbol{\theta}$ and $\boldsymbol{\eta}$. If $\boldsymbol{o_n}$ is the vector to which the transformation is applied, the output of the linear FC layer is twice the length of $\boldsymbol{o_n}$. Consequently, FiLM performs an affine transform on $\boldsymbol{o_n}$ using $\boldsymbol{\theta}$ and $\boldsymbol{\eta}$:
\begin{align}
        \boldsymbol{q} &= \boldsymbol{\theta} \boldsymbol{o_n} + \boldsymbol{\eta}.
\end{align}

Following the projection, it is common to apply a function, such as ReLU or sigmoid, to determine the amount of information that should be passed. In our approach, we replace the activation function with a GLU layer. This layer is more stable than ReLU and exhibits faster learning compared to sigmoid. Similarly to the previous step, the GLU layer consists of a linear FC layer that takes the FiLM output vector as input and computes a vector twice its length. The resulting output is then divided equally into two vectors: $\boldsymbol{q_1}$ and $\boldsymbol{q_2}$. A function is applied to $\boldsymbol{q_2}$, with the resulting output being multiplied element-wise with $\boldsymbol{q_1}$. This operation describes the conditioning block, represented by the following equation:
\begin{align}
     \boldsymbol{o^c_n} = \boldsymbol{q_1} \odot softsign(\boldsymbol{q_2})
\end{align}
The GLU layer determines the flow of information through the network, acting as a logical gate. Typically, a sigmoid activation function is used for this purpose. However, in our approach, we introduce a softsign function instead. The softsign function controls the extent to which the control parameters should positively or negatively influence the final output. A value of $0$ indicates a bypass, meaning no influence on the output. This allows for more flexibility.

Lastly, the presented models are designed to have limited computational complexity for the inference process. Specifically, we set the limit to $1.500$ FLOPs per sample or $72$ MFLOPs per second at $48$ kHz. By doing so, the models feature $\sim 750$ internal parameters. Table~\ref{tab:flops} reports the FLOPs per sample for each model.
\begin{table}[h]
\caption{FLOPs per sample for each model considered in this study.}\label{tab:flops}
\begin{tabular*}{\textwidth}{lcccc}
\toprule%
Model & Recurrent Layer & Conditioning Block & Total\\
\midrule
LSTM & $1040$ & $120$ & $1160$\\
ED & $928$ & $120$ & $1048$\\
LRU & $692$ & $120$ & $812$\\
S4D & $792$ & $120$ & $912$\\
S6 & $864$ & $120$ & $984$\\

\end{tabular*}
\end{table}

\subsection{Losses and Metrics}\label{sec:metrics}

The models in this study take raw audio samples as input and produce raw audio samples as output. The most common losses used in this case include the Mean Squared Error (MSE), the Mean Absolute Error (MAE), and the Error-to-Signal Ratio (ESR). Based on our knowledge and supported by empirical experiments, there is no evidence suggesting that more complex loss functions could benefit the learning process, especially when modeling the transformation applied to a signal by a generic audio effect. While using specific and different loss functions for different effects may potentially yield better results, we decided to carry out this study using a fixed loss function across architectures and audio effects. This allows for a fair comparison between the trained models, as they all attempt to minimize the same loss function. Therefore, we selected the MSE for this comparative study as it helps to capture the dynamic aspects of the output signal more effectively ~\cite{simionato2023fully}.

In addition, our quantitative evaluation is based on various metrics to assess the models from different perspectives.  In the following expressions, $\boldsymbol{y}$ and $\hat{\boldsymbol{y}}$ represent the target and predicted audio signals, and the symbol $N$ represents the overall number of samples in the output signal.

The first metric we utilize is the normalized root-mean-square energy (NRMSE) error, with normalization carried out using the RMSE of the target signal. NRMSE error provides insights into the energy deviation between the target and the prediction.

The losses and metrics presented so far calculate the average difference between the target and prediction, which may hide large errors that occur only during signal transients. Transients, such as sharp attacks or decays, may represent a small fraction of the overall duration of an audio signal. In this case, we may obtain a relatively low error even with all transients poorly modeled. Therefore, we have included a metric related to the spectral flux, which considers the differences between pairs of spectra computed on consecutive overlapping windows, as used in the library~\footnote{\url{https://github.com/magenta/ddsp}} developed after ~\cite{engel2020ddsp}:
\begin{equation}\label{eq:diff}
\begin{split}
  M_{SF} = \Bigg|\Bigg| \frac{ \Big( \Big| |\mathrm{STFT}(\boldsymbol{y})_n| - |\mathrm{STFT}(\boldsymbol{y})_{n-1}| \Big| \Big)  
    - \Big(\Big| |\mathrm{STFT}(\hat{\boldsymbol{y}})_n| - |\mathrm{STFT}(\hat{\boldsymbol{y}})_{n-1}| \Big| \Big) }{\Big| |\mathrm{STFT}(\boldsymbol{y})_n| - |\mathrm{STFT}(\boldsymbol{y})_{n-1}| \Big|} \Bigg|\Bigg|_1
    \end{split}
\end{equation}
where the Short-Time Fourier Transform (STFT) is computed with a window size equal to $2048$ and hop size equal to $512$ samples. The metric is normalized by the squared difference between the magnitudes of the spectra of two successive target frames, measuring the spectral change between them. This indicates the rate at which the power spectrum of a signal is evolving over time. 

To account for the frequency response, we use the multi-resolution STFT ~\cite{engel2020ddsp}:
\begin{equation}\label{eq:stft}
\begin{split}
    M_{STFT} &= \frac{1}{N} \sum_m \Big|\Big| \frac{ |\mathrm{STFT}_m (\boldsymbol{y})| - |\mathrm{STFT}_m (\hat{\boldsymbol{y}})| }{  |\mathrm{STFT}_m (\boldsymbol{y})| } \Big|\Big|_1\\
    &+ \frac{1}{N} \sum_m|| \log(|\mathrm{STFT}_m (\boldsymbol{y})|) - \log(|\mathrm{STFT}_m (\hat{\boldsymbol{y}})|) ||_1 
    \end{split}
\end{equation}
which compares linear and logarithmic spectral distances with varying frequency resolutions using the L1 norm. The linear component of $M_{STFT}$ is normalized by the magnitude of the STFT of the target. The error is calculated as the sum of the absolute values of the vectors. This metric quantifies the error between the spectra employing different resolutions. We compute the metrics using $m = [256, 512, 1024]$ as resolutions.

\subsection{Audio Effects and Datasets}\label{sec:datasets}

In this section, we introduce the audio effects selected for the study and the composition of their datasets. We proceeded to record the dataset using effects that were at our disposal. In particular, we selected the following effects: overdrive, saturation, equalization, low-pass filter, and compression. Table \ref{tab:datasets} summarizes all the datasets included in the study. Delay-based effects require a different approach than a direct black box ~\cite{wright2021neural}, ~\cite{mikkonen2023neural}. For this reason, they are not included in this study.

The dataset we use is derived from the following hardware devices: Behringer OD300 overdrive pedal~\footnote{\url{https://www.behringer.com/product.html?modelCode=P0608}}, Behringer Neutron~\footnote{\url{https://www.behringer.com/product.html?modelCode=P0CM5}}'s overdrive module,  Behringer Neutron's filter module in low-pass mode,
TC Electronic Bucket Brigade Analog Delay used as a saturator~\footnote{\url{https://www.tcelectronic.com/product.html?modelCode=P0EBV}}, CL 1B TubeTech~\footnote{\url{http://www.tube-tech.com/cl-1b-opto-compressor/}}and Teletronix LA-2A~\footnote{\url{https://www.uaudio.com/hardware/la-2a.html}} optical compressors. We have also collected data from two software VA plugins using ~\cite{fasciani2024auniversal}: the Helper Saturator~\footnote{\url{https://www.waproduction.com/plugins/view/helper-saturator}} and Universal Audio Pultec EQ~\footnote{\url{https://www.uaudio.com/uad-plugins/equalizers/pultec-passive-eq-collection.html}}, because the hardware saturator presents no variable control parameter, and because we have no analog equalizer in our immediate availability.

\paragraph{Overdrive} Both the overdrive devices feature the following control parameters: volume, distortion level, and tone. The OD300 is a distortion pedal that features two different types of distortion, labeled overdrive and distortion modes. To record the dataset, the unit was set to overdrive mode. The device was sampled at $5$ different and equally spaced values for the distortion and tone knobs.
Neutron's overdrive is a module integrated into the Neutron synthesizer. Thanks to the semi-modular nature of this device, we were able to isolate the overdrive module. We recorded the dataset's varying distortion and tone as we did for the OD300, but sampling $10$ equally spaced values for each parameter.
In both devices, the volume level controls an attenuator, and we fixed this parameter to the maximum (i.e., no attenuation).

\paragraph{Filter}

Similarly to the overdrive, the Behringer Neutron integrates a filter module, which features frequency cutoff and resonance knobs. The filter can be set as low-pass, high-pass, and band-pass. We isolated the module and recorded the dataset at $10$ equally spaced values for the cutoff and $5$ for the resonance, selecting the low-pass filter type. The filter was separated from the envelope module, which can be used to modulate it.

\paragraph{Saturator}

The Helper Saturator software emulates analog saturation, in particular tube and tape saturation. The software presents four parameters: lowpass and highpass cutoffs, saturation type switch (tape or tube), and saturation level. We recorded the dataset for $10$ different values of the saturation level. The saturation switch was set to tape mode. Lowpass and highpass cutoffs were set to $20$ and $20000$ Hz, respectively.

\paragraph{Equalizer}

The software plug-in Universal Audio Pultec Passive Equalizer emulates the tube-based equalizer Pultec EQP-1A. The effect presents low-frequency ([20, 30, 60, 100] Hz) and high-frequency ([3, 4, 5, 8, 10, 12, 16]kHz) switches to select the frequencies to boost or cut using the respective boost and attenuator knobs ([0,10]). Additionally, the emulation also has a bandwidth knob ([0,10]). The equalizer boosts and cuts the selected frequency bands, giving some saturation simultaneously. In this case, we recorded the dataset varying the boost and attenuator knobs at $5$ different equally-spaced values up to half of their range, setting $60$~Hz as the low-frequency switch value and $0$ (sharp) as the bandwidth.

\paragraph{Compressor/Limiter}

The CL 1B and the LA-2A are analog optical compressors. In optical compressors, a lighting-emitting element is fed with the audio signal that illuminates a light-sensitive resistor. The input signal's amplitude determines the element's brightness, which, in turn, changes the resistance in the gain attenuation circuit. We recorded the dataset directly from the compressor. The CL 1B compressor dataset presents four parameters: attack, release, ratio, and threshold. We consider three equally spaced values for the ratio, attack, and release time and four values for the threshold. On the other hand, LA-2A presents the peak reduction knob indicating the amount of compression ([$0$,$100$]) and a mode switch (limiter, compression). We consider $10$ equally spaced values for the peak reduction.

\begin{table}[h]
\caption{Datasets with related parameters and ranges considered in the study.}\label{tab:datasets}
\begin{tabular*}{\textwidth}{lcccccc}
\toprule%
  & Type & Parameters & Range & Combinations \\
\toprule%
OD300\footnotemark[1] & Overdrive & Level & [min, max] & $25$\\
 &  & Tone & [min, max]  \\
 \midrule
Neutron & Overdrive & Level & [min, max]  & $100$ \\
Overdrive module\footnotemark[1] &  & Tone & [min, max] \\
 \midrule
Neutron  & Low pass  & Cutoff & [min, max] & $100$ \\
Filter module\footnotemark[1] & filter & Resonance & [min, max] \\
 \midrule
Helper  & Saturator &  Saturation & [min, max] & $10$\\
Saturator\footnotemark[2] &  &  Saturation Type & Tape \\
  \midrule
Pultec  & Equalizer & Low frequency & 60 Hz & $25$  \\
Passive EQ \footnotemark[2] &   & Frequency booster & [0,10]  \\
 &   & Bandwidth & 0 \\
 &   & Frequency Attenuator & [0,10] \\
\midrule
LA-2A\footnotemark[1] & Optical  & Peak Reduction & [0,100] & $20$ \\
  & Compressor & Switch Mode  & [Compressor,Limiter]  \\
  \midrule
CL 1B\footnotemark[1] & Optical  & Threshold & [-40, 0] dB & $108$ \\
 &  Compressor & Ratio & 1:[1, 10]  \\
 &  & Attack & [5, 300] ms  \\
 &  & Release & [0.005, 10] s \\

\end{tabular*}
\footnotetext[1]{Recorded from hardware devices.}
\footnotetext[2]{Recorded from software devices.}
%\footnotetext[3]{Publicly available.}
\end{table}

\subsubsection{Data Collection}

The data collection process was carried out using a MOTU M4 audio interface to feed a selection of audio signals into the system and simultaneously record its output. To achieve this, the left input channel of the audio interface was connected to the left output channel of the interface itself. The right input channel of the audio interface was connected to the output of the device, and the input of the device to the right output channel of the audio interface. This allows effective recording of both the device’s input and output signals, compensating for the minor sound coloring and latency of the audio interface. The audio data was recorded at a sampling rate of $48$ kHz.

The input mono signal has a duration of $45$ seconds for each parameter combination and includes a variety of sounds, such as frequency sweeps covering a range of $20$ Hz to $20$ kHz, white noises with increasing amplitudes (both linear and logarithmic), recordings of instruments such as guitar, bass, drums (both loops and single notes), vocals, piano, pad sounds, and sections from various electronic and rock songs. Finally, the control parameter of each effect was mapped to the range of [0,1].

\subsection{Experimenting and Learning}

The models are trained for $200$ epochs and use the Adam \cite{kingma2014adam} optimizer with a gradient norm clipped at $1$ ~\cite{pascanu2013difficulty}. The training was stopped earlier in case of no reduction of validation loss for $10$ epochs. We design a learning rate with an exponential decay according to the formula: $lr = LR*0.25^{e}$, where LR is the initial learning rate and it is set to $3 \cdot 10^{-4}$, and $e$ is the number of epochs. Test losses and evaluation metrics are computed using the model’s weights that minimize the validation loss throughout the training epochs. Finally, the input signal is split into segments of $2400$ samples (equivalent to $50$ ms) to be processed before updating the weights. Considering the network architecture described in Section~\ref{sec:nn}. Consequently, our implementation adopts an input shape of ($B$, $L$, $F$), which is set to ($B$, $2400$, $64$), where $B$ denotes the arbitrary batch size, $L$ the sequence length, and $F$ the input dimensionality, often identified as the number of input features. This configuration is used within the framework of Truncated Backpropagation Through Time (TBPTT).

We adopted the same learning schedule and minibatch size for all models across all datasets. As explained in Section~\ref{sec:datasets}, we collected the datasets using the same approach and the same input file to minimize the differences among them. However, it is important to highlight that each dataset and modeling task may benefit from specific adjustments to achieve more accurate results. Different datasets and tasks may require slightly different learning rate schedules; an excessively slow or fast learning rate can lead to suboptimal solutions. The number of samples to process before updating the weights, referred to as the minibatch size, impacts the frequency of updating weights: larger sizes may result in faster convergence but poorer generalization, while smaller sizes can aid generalization but may require additional epochs.

To ensure robust evaluation of the models, we divided the collected data into $5$ distinct compositions of training, validation, and test sets. This strategy aimed to cross-validate the models across different data splits. In each composition, $80\%$ of the data is allocated for training, $10\%$ is reserved for validation, and the remaining $10\%$ constitutes the test set. Importantly, the validation and test sets in each composition include different data signals, ensuring that the model never uses the same audio signals in both validation and testing processes. This approach prevents any bias towards certain data signals and allows for a more accurate assessment of the model's performance across varying conditions and inputs.

For audio effects with variable control parameters, each $45$ s recording in the dataset corresponds to a specific combination of effect parameters. To maintain consistency across training, validation, and testing, we employ a $80-10-10\%$ split at the level of individual recordings. This ensures that all combinations of control parameters are proportionally represented in each subset, providing balanced usage for both training the model and evaluating its performance. Additionally, minor manual adjustments were made to the splitting points to ensure they occur during silent segments, preventing abrupt changes in audio dynamics that could affect model accuracy.

The statistical significance of the results is analyzed using the Friedman test and the Wilcoxon signed-rank test ~\cite{demvsar2006statistical, rainio2024evaluation}. Both are non-parametric statistical tests employed to determine whether there are statistically significant differences between the medians of samples. These tests are well-suited for data that does not meet the assumptions required for parametric tests, such as normality.

In this case, where we are evaluating $5$ models across $5$ different dataset compositions, the Friedman test is the most appropriate for assessing statistical significance. However, due to the limited number of dataset compositions, the reliability of the Friedman test might be reduced. To address this, we also conduct separate Wilcoxon signed-rank tests for each model pair. While this test is useful for assessing significant differences between two models, it's important to note that it is not ideal for comparing multiple models. Specifically, conducting multiple Wilcoxon tests increases the risk of incorrectly finding significant differences due to the numerous comparisons involved.

\section{Results}\label{sec:results}

The MSE loss, the metrics described in Section \ref{sec:metrics}, and the mean of the epochs resulting in lower validation loss during training are reported in Table~\ref{tab:errors}. The losses and the metrics consider all conditioning scenarios and provide an average error across all combinations of parameters. The average epoch at which the training met the early stopping condition is also reported in the table. \textbf{LRU} is generally the slowest to converge, taking significantly more epochs to reach the same range of errors as the other models. On the other hand, other variants are dataset-specific, with a tendency for the SSM-based model, such as \textbf{S4D} and \textbf{S6}, to converge faster and with a similar number of epochs. Similarly, both \textbf{LSTM} and \textbf{ED} require a comparable number of epochs, although they generally converge more slowly than models based on SSM. This slower convergence is attributed to their similar layer formulation. On the other hand, \textbf{LRU} utilizes specific initializations and normalization schemes, distinguishing it from traditional recurrent and SSM architectures.
\begin{table}[h!]
\caption{Mean values of MSE loss, ESR, $M_{NRMSE}$, $M_{SF}$, and $M_{STFT}$ metrics computed across different test sets, with each set linked to a model trained on a distinct dataset split. The last column reports the average epoch at which the training met the early stopping condition.}\label{tab:errors}
\footnotesize
\begin{tabular*}{\textwidth}{llcccccl}
\toprule%
Dataset & Model & MSE & ESR & $M_{NRMSE}$ & $M_{SF}$ & $M_{STFT}$ & Epoch\\
\midrule
\midrule
OD300 & LSTM & $\boldsymbol{2.92\cdot 10^{-4}}$ 
& $\boldsymbol{1.13 \cdot 10^{-1}}$ 
& $\boldsymbol{2.69 \cdot 10^{-1}}$ 
& $\boldsymbol{4.56\cdot 10^{-1}}$ 
&  $ 7.50\cdot 10^{-1}$ 
& $157$
\\
Overdrive & ED  & $ 3.59\cdot 10^{-4}$ 
& $2.00\cdot 10^{-1} $ 
& $2.90 \cdot 10^{-1} $ 
& $ 5.66\cdot 10^{-1} $ 
& $ 7.78\cdot 10^{-1} $ 
& $158$
\\
& LRU &  $ 4.57\cdot 10^{-4} $
 & $1.77\cdot 10^{-1}  $
 & $3.44\cdot 10^{-1}  $
 & $5.35\cdot 10^{-1}  $
 & $8.12\cdot 10^{-1}  $
 & $170$
\\
& S4D & $3.14 \cdot 10^{-4} $
 & $1.22\cdot 10^{-1}  $
 & $2.88\cdot 10^{-1}  $
 & $4.89 \cdot 10^{-1} $
 & $\boldsymbol{7.40\cdot 10^{-1}}$
& $189$
\\
& S6  & $4.45 \cdot 10^{-4} $
 & $1.72\cdot 10^{-1} $
 & $3.57\cdot 10^{-1}  $
 & $5.43\cdot 10^{-1}  $
 & $8.82\cdot 10^{-1} $
 & $153$
 \\
\midrule
\midrule
Neutron & LSTM & $\boldsymbol{5.64\cdot 10^{-4}}$ 
& $\boldsymbol{6.21 \cdot 10^{-2}}$ 
& $\boldsymbol{2.51 \cdot 10^{-1}}$ 
& $\boldsymbol{3.25 \cdot 10^{-1}}$ 
& $5.45 \cdot 10^{-1}$ 
& $18$
\\
Overdrive & ED  & $7.84 \cdot 10^{-4} $
 & $8.64\cdot 10^{-2} $
 & $2.87\cdot 10^{-1}  $
 & $3.53\cdot 10^{-1}  $
 & $\boldsymbol{5.42\cdot 10^{-1}}$
 & $54$
 \\
& LRU  & $9.65\cdot 10^{-4} $
 & $1.10 \cdot 10^{-1} $
 & $3.10\cdot 10^{-1}  $
 & $4.09 \cdot 10^{-1} $
 & $5.98 \cdot 10^{-1} $
& $106$
\\
& S4D  & $1.12 \cdot 10^{-3} $
 & $1.30\cdot 10^{-1} $
 & $4.20\cdot 10^{-1}  $
 & $5.33 \cdot 10^{-1} $
 & $5.98 \cdot 10^{-1} $
 & $10$
 \\
& S6  & $1.16\cdot 10^{-3}$
 & $1.32 \cdot 10^{-1}$
 & $4.42\cdot 10^{-1}$
 & $5.08 \cdot 10^{-1} $
 & $6.58 \cdot 10^{-1}$
  & $10$
\\
\midrule
\midrule
Helper & LSTM & $3.47 \cdot 10^{-4} $ 
& $ 1.91\cdot 10^{-1}  $ 
& $3.81 \cdot 10^{-1} $ 
& $\boldsymbol{4.65\cdot 10^{-1}}$ 
& $6.88 \cdot 10^{-1} $ 
& $180$
\\
Saturator & ED & $3.49 \cdot 10^{-4} $
 & $1.66 \cdot 10^{-1} $
 & $3.82\cdot 10^{-1} $
 & $4.82 \cdot 10^{-1} $
 & $6.70 \cdot 10^{-1} $
 & $94$
\\
& LRU  & $6.11 \cdot 10^{-4} $
 & $3.36 \cdot 10^{-1}  $
 & $4.96 \cdot 10^{-1} $
 & $5.54 \cdot 10^{-1} $
 & $8.90 \cdot 10^{-1} $
 & $157$
\\
& S4D & $\boldsymbol{2.93\cdot 10^{-4}}  $
 & $\boldsymbol{1.62 \cdot 10^{-1}} $
 & $\boldsymbol{3.54\cdot 10^{-3}} $
 & $4.72 \cdot 10^{-1} $
 & $6.59\cdot 10^{-1} $
 & $109$
 \\
& S6  & $2.99 \cdot 10^{-4}$
 & $1.63\cdot 10^{-1}$
 & $3.57\cdot 10^{-1}$
 & $4.85 \cdot 10^{-1}$
 & $\boldsymbol{6.57\cdot 10^{-1}}$
 & $176$
\\
\midrule
\midrule
Pultec & LSTM & $\boldsymbol{ 4.26 \cdot 10^{-5} }$
 & $\boldsymbol{1.14\cdot 10^{-2}}$
 & $\boldsymbol{9.63 \cdot 10^{-3}}$
 & $\boldsymbol{7.11\cdot 10^{-2}}  $
 & $\boldsymbol{1.85 \cdot 10^{-1}}$ 
 & $66$\\
Passive & ED &  $ 1.43\cdot 10^{-4}$
 & $ 3.49\cdot 10^{-2}  $
 & $ 1.56\cdot 10^{-1}  $
 & $ 1.18\cdot 10^{-1}  $
 & $2.92 \cdot 10^{-1} $
& $130$ \\
Equalizer & LRU  & $ 2.19\cdot 10^{-4}$
 & $ 5.60\cdot 10^{-2}  $
 & $ 1.98\cdot 10^{-1}  $
 & $ 2.17\cdot 10^{-1}  $
 & $3.62\cdot 10^{-1} $ 
 & $173$
 \\
& S4D  & $1.62\cdot 10^{-4}  $
 & $6.97\cdot 10^{-2}  $
 & $8.73\cdot 10^{-2}$
 & $7.46\cdot 10^{-2} $
 & $2.28\cdot 10^{-1}  $
  & $8$
  \\
& S6  & $4.79\cdot 10^{-5}  $
 & $ 1.20\cdot 10^{-2}  $
 & $ 1.00\cdot 10^{-1}  $
 & $ 8.41 \cdot 10^{-1} $
 & $ 1.96 \cdot 10^{-1} $
& $56$
\\
\midrule
\midrule
LA-2A & LSTM & $1.90\cdot 10^{-4}$ 
& $1.11\cdot 10^{-1}$
& $3.11\cdot 10^{-1}$ 
& $3.34\cdot 10^{-1}$ 
& $6.61\cdot 10^{-1}$
& $57$
\\
Optical & ED &  $ 4.22\cdot 10^{-5}$ 
& $2.36\cdot 10^{-2}$ 
& $1.60\cdot 10^{-1}$
& $1.78 \cdot 10^{-1}$ 
& $3.47 \cdot 10^{-1}$
& $70$ 
\\
Comp & LRU & $5.08\cdot 10^{-5}$ 
& $2.82\cdot 10^{-2}$ 
& $1.67\cdot 10^{-1}$ 
& $2.22\cdot 10^{-1}$ 
& $3.97\cdot 10^{-1}$
 & $154$
\\
& S4D  & $ 3.76\cdot 10^{-5}$ 
& $2.07\cdot 10^{-2}$
& $1.47\cdot 10^{-1}$ 
& $1.62\cdot 10^{-1}$ 
& $3.44 \cdot 10^{-1}$
& $54$
\\
& S6 & $\boldsymbol{3.60\cdot 10^{-5}}$ 
& $\boldsymbol{1.96\cdot 10^{-2}}$
& $\boldsymbol{1.43\cdot 10^{-1}}$ 
& $\boldsymbol{1.53 \cdot 10^{-1}}$ 
& $\boldsymbol{3.31\cdot 10^{-1}}$
& $22$
\\
\midrule
\midrule
CL 1B & LSTM & $3.37\cdot 10^{-4} $ 
& $2.72\cdot 10^{-2}$ 
& $4.88\cdot 10^{-1}$ 
& $5.97\cdot 10^{-1}$ 
& $9.16\cdot 10^{-1}$
& $35$
\\
Optical & ED & $\boldsymbol{2.68\cdot 10^{-5}}$ 
& $\boldsymbol{2.01\cdot 10^{-2}}$ 
& $\boldsymbol{1.20\cdot 10^{-1}}$ 
& $\boldsymbol{1.43\cdot 10^{-1}}$ 
& $\boldsymbol{3.51 \cdot 10^{-1}}$
& $52$ 
\\ 
Comp & LRU & $6.21\cdot 10^{-5}$ 
& $5.02\cdot 10^{-2}$ 
& $1.96\cdot 10^{-1}$ 
& $2.23 \cdot 10^{-1}$ 
& $4.59\cdot 10^{-1}$
& $168$
\\
& S4D & $1.09\cdot 10^{-4}$ 
& $9.06\cdot 10^{-2}$ 
& $2.78\cdot 10^{-1}$ 
& $2.95\cdot 10^{-1}$ 
& $5.83 \cdot 10^{-1}$ 
& $28$
\\
& S6 & $ 6.64\cdot 10^{-5}$ 
& $5.12\cdot 10^{-2}$
& $1.83\cdot 10^{-1}$ 
& $2.00\cdot 10^{-1}$ 
& $4.66 \cdot 10^{-1}$
& $14$
\\
\midrule
\midrule
Neutron & LSTM & $3.18\cdot 10^{-3}$ 
& $7.03 \cdot 10^{-1}$ 
& $8.28\cdot 10^{-1}$ 
& $8.62\cdot 10^{-1}$
& $1.19$ 
& $76$\\
Filter & ED  & $2.91\cdot 10^{-3}$ 
& $6.38\cdot 10^{-1}$ 
& $\boldsymbol{6.82\cdot 10^{-1}}$ 
& $\boldsymbol{7.57 \cdot 10^{-1}}$ 
& $1.13$
& $54$
\\
& LRU &  $2.62 \cdot 10^{-3}$ 
& $ 5.77 \cdot 10^{-1} $
& $7.18\cdot 10^{-1}$ 
& $8.01 \cdot 10^{-1} $ 
& $1.08$
& $179$
\\
& S4D  & $\boldsymbol{2.52\cdot 10^{-3}}$ 
& $\boldsymbol{5.56\cdot 10^{-1}}$ 
& $ 6.94\cdot 10^{-1} $ 
& $7.90 \cdot 10^{-1} $ 
& $1.06$
& $16$\\
& S6  & $ 2.62\cdot 10^{-3}$ 
& $ 5.77\cdot 10^{-1}$ 
& $ 7.21\cdot 10^{-1}$ 
& $ 8.11\cdot 10^{-1}$  
& $\boldsymbol{1.04}$
& $7$
\\
\end{tabular*}
\end{table}
\begin{figure}[h!]
    \begin{center}
    \includegraphics[scale=0.39]{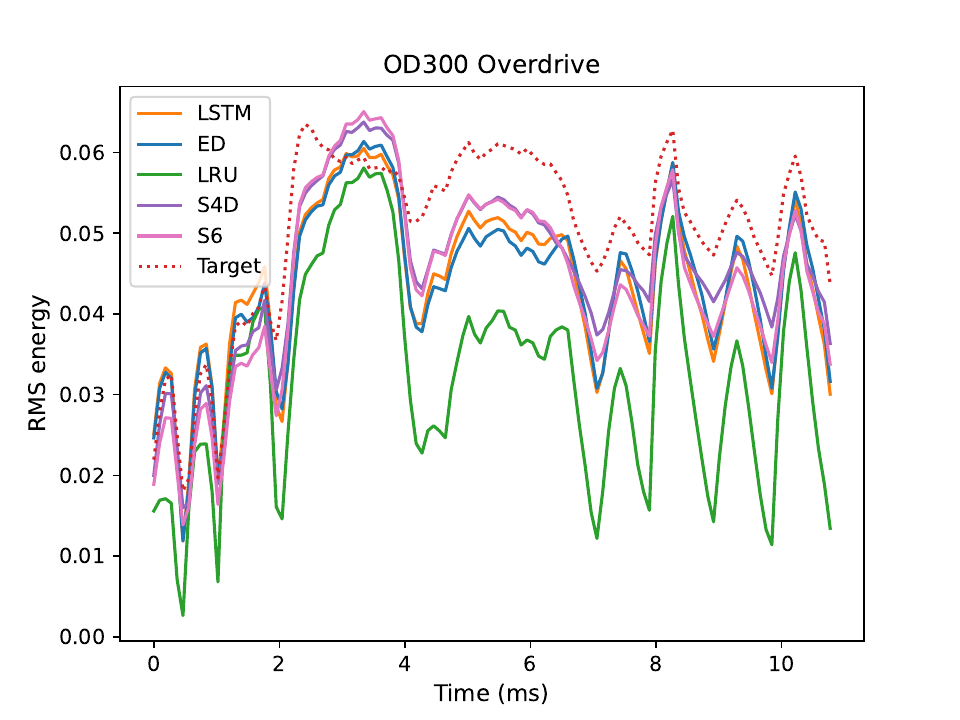}
    \includegraphics[scale=0.39]{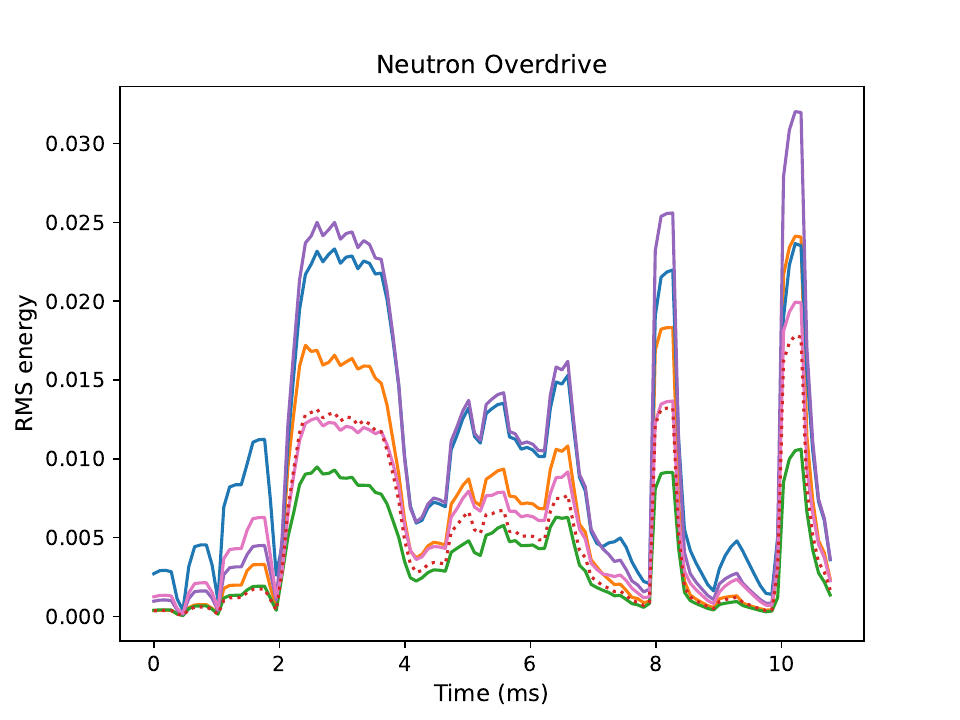}\\
    \includegraphics[scale=0.39]{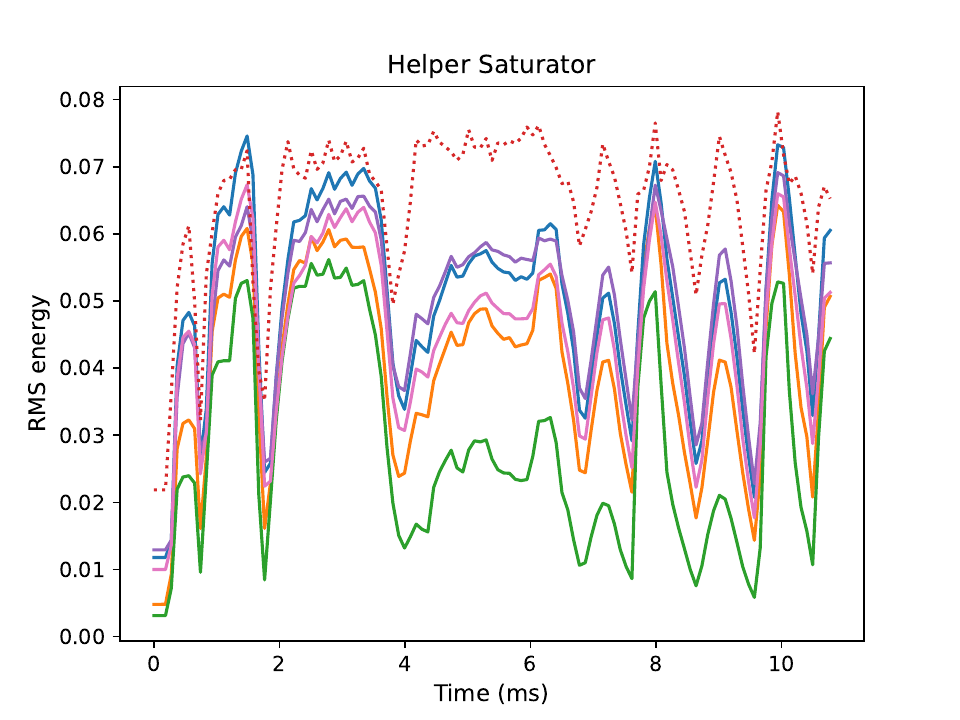}
    \includegraphics[scale=0.39]{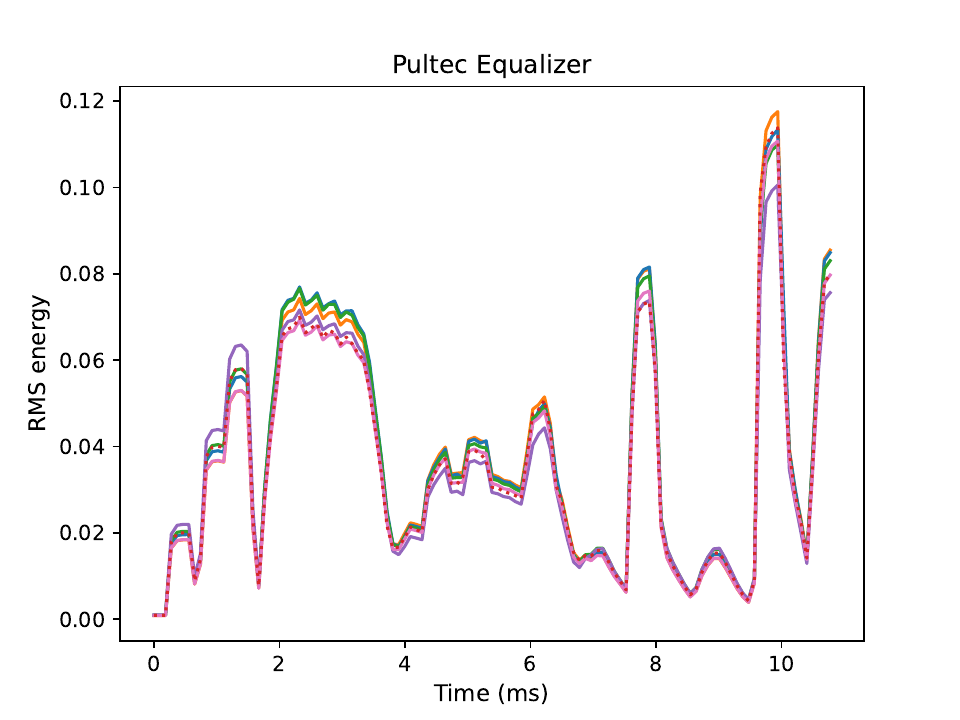}\\
    \includegraphics[scale=0.39]{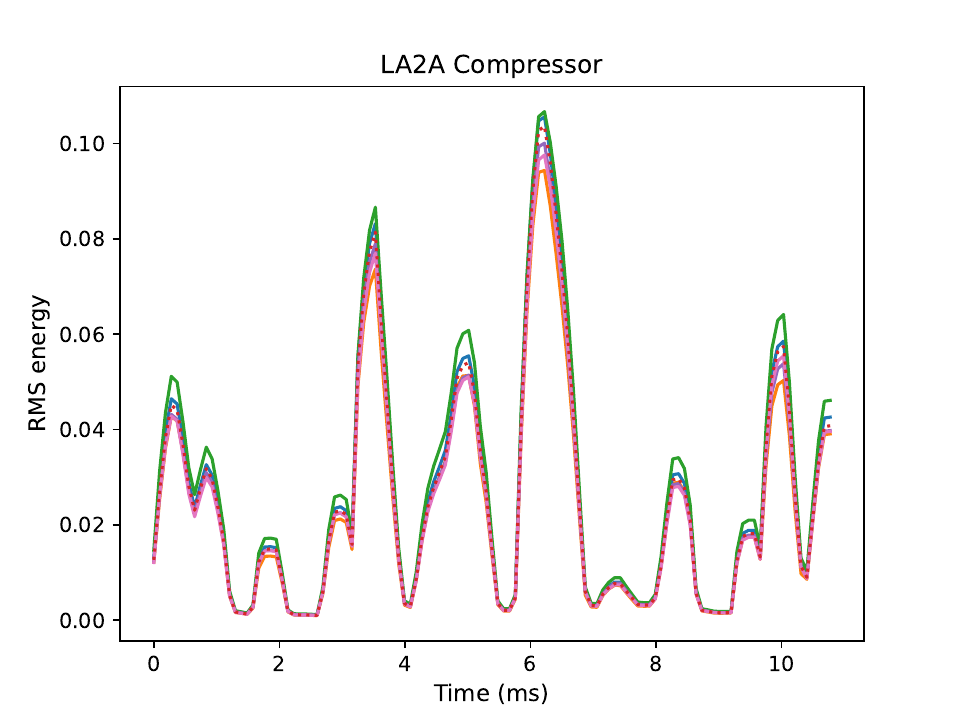}
    \includegraphics[scale=0.39]{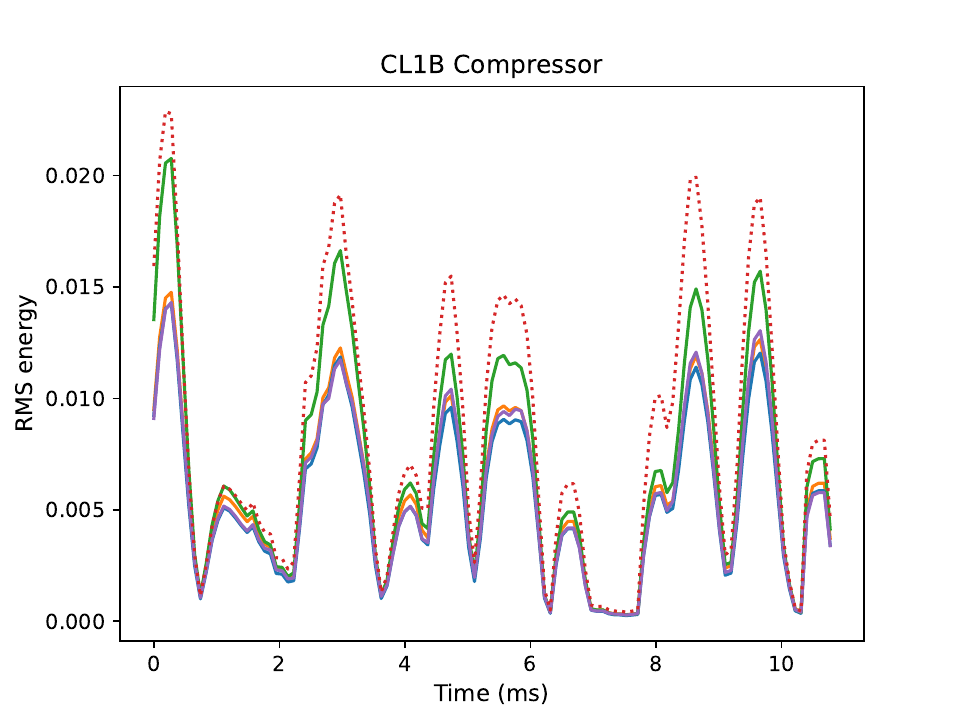}\\
     \includegraphics[scale=0.39]{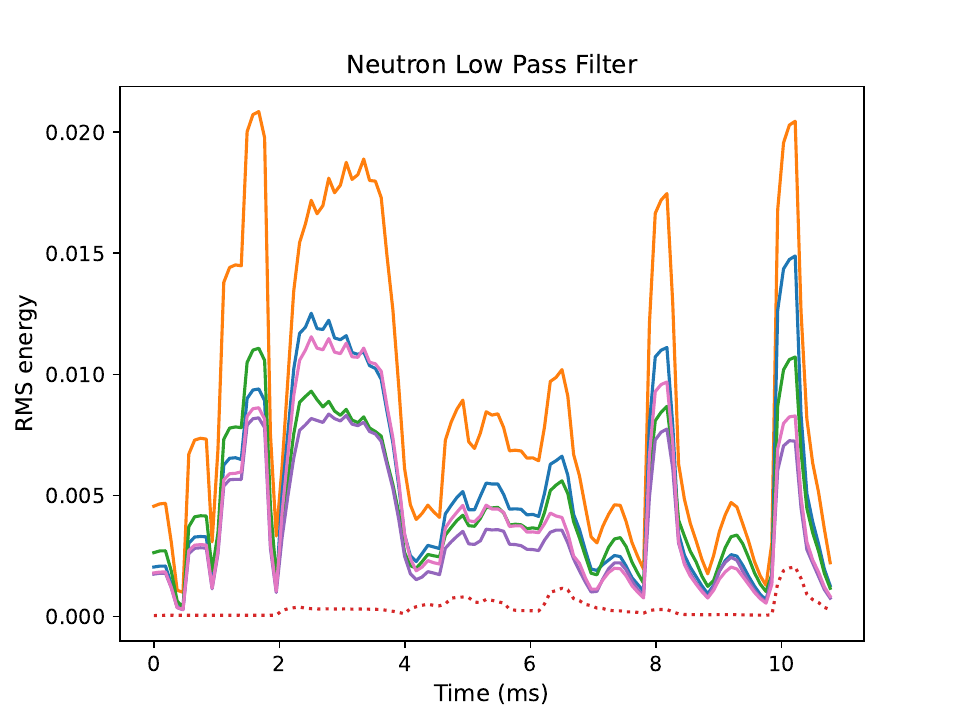}\\
    \end{center}
     \caption{Comparison of the RMS energy between the target and predicted output. The RMS is computed using windows of $4096$ samples with $75\%$ overlap. When the model includes conditioning, the results refer to the highest conditioning values, representing the most challenging scenario.}\label{fig:rmse}
\end{figure}
In the OD300 and Neutron's overdrive dataset, the \textbf{LSTM} model demonstrates superior overall performance. However, it does not achieve the lowest $M_{STFT}$, although its values are close to the best. Additionally, \textbf{S6} struggles more with emulating both effects, particularly with the Neutron's overdrive dataset. When trained on the Saturator dataset, \textbf{S4D}, \textbf{S6}, and \textbf{LRU} deliver the strongest performance, with \textbf{LSTM} and \textbf{ED} slightly trailing. In contrast, for the Equalizer dataset, \textbf{LSTM} consistently performs better than the other models.

The optical compressor features extended transients and lengthy time dependencies between the input signal and the alteration applied by the device, influenced by its adjustable attack and release parameters. This complexity is especially pronounced with the CL 1B compressor, presenting a challenging task for ANN modeling. Of the tested models, \textbf{LSTM} is the least accurate, with higher losses in both the LA-2A and CL 1B scenarios. In contrast, \textbf{ED}, \textbf{S4D}, and \textbf{S6} demonstrate superior accuracy for these tasks, with \textbf{S6} excelling in the LA-2A case and \textbf{ED} in the CL 1B case.

In analyzing the metrics of models trained to emulate the low-pass filter, both \textbf{S4D} and \textbf{ED} show the best performance. However, all models reveal a considerable frequency mismatch with the target output. This suggests their inability to model the signal alteration process of this parametric effect, which selectively removes high-frequency components while leaving those below the cutoff frequency unchanged.

Overdrive effects significantly distort the input signal, leading to the introduction of harmonic components. In these cases, LSTM-based models have been shown to be more effective. Conversely, the saturator determines mild distortion, similar to tape processing, by smoothing out the transients of the input signal. This situation may involve hysteresis and short delays associated with magnetic recording ~\cite{mikkonen2023neural}, leading to longer temporal dependencies between input and output compared to distortion effects. This characteristic may explain the advantages observed with SSM-based layers. However, we lack information about the specific implementation of the software used, preventing us from confirming this hypothesis. Conversely, The Equalizer introduces tube-like saturation, which may explain the better performance of LSTMs in this context. SSM-based layers also demonstrated improved performance with optical compressors, where the effective management of transients and temporal dependencies is crucial.

When ranging the models according to performance metrics, inconsistencies arise with $M_{SF}$ and $M_{STFT}$, as these metrics don't always align with the trends observed in time-based metrics. Nonetheless, no distinct patterns are apparent. This could be due to the loss function's emphasis on the time domain, where variations in values are typically subtle.

\begin{figure}[h!]
    \begin{center}
    \includegraphics[scale=0.39]{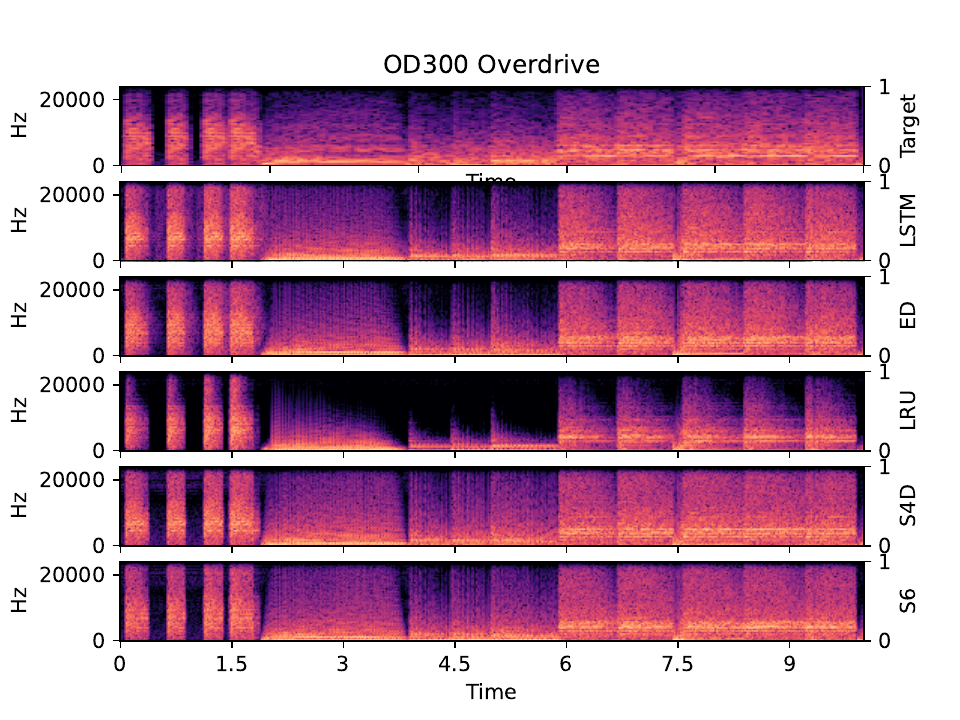}
    \includegraphics[scale=0.39]{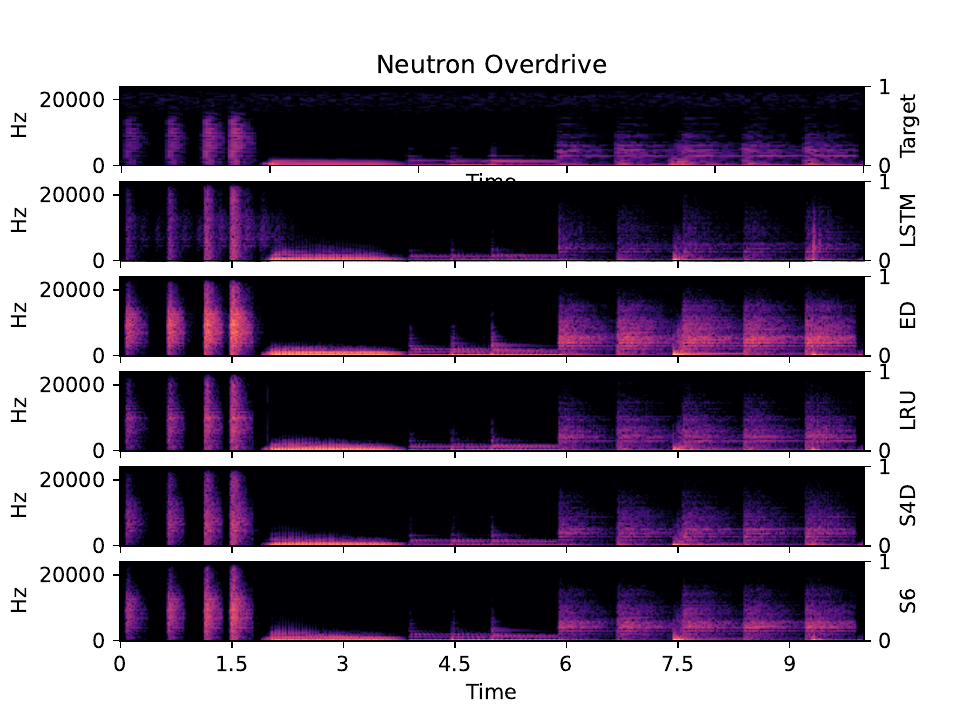}\\
    \includegraphics[scale=0.39]{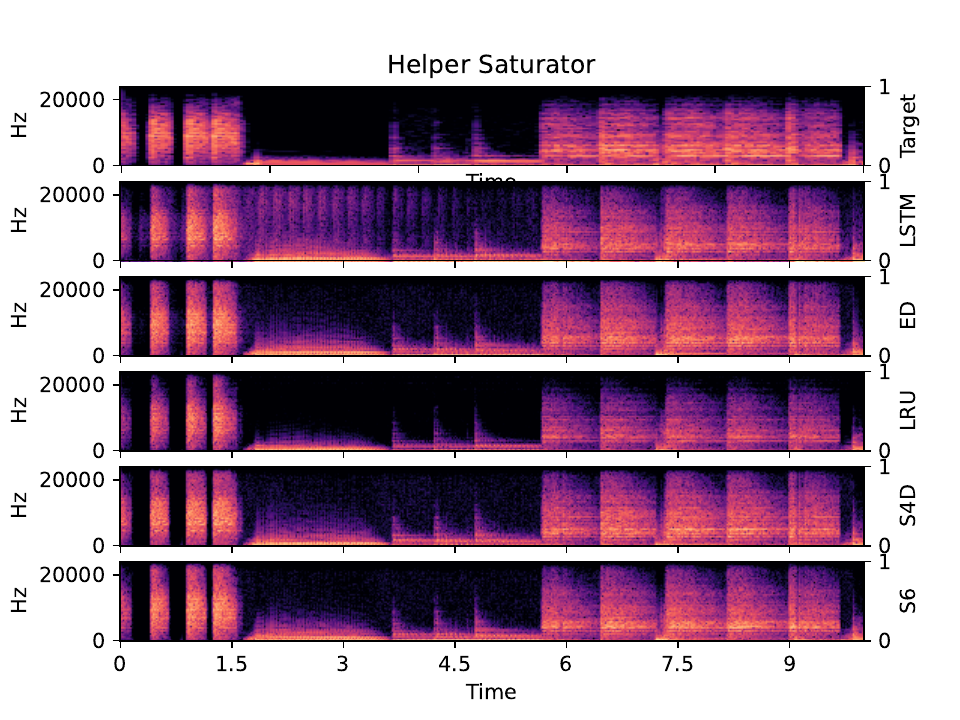}
    \includegraphics[scale=0.39]{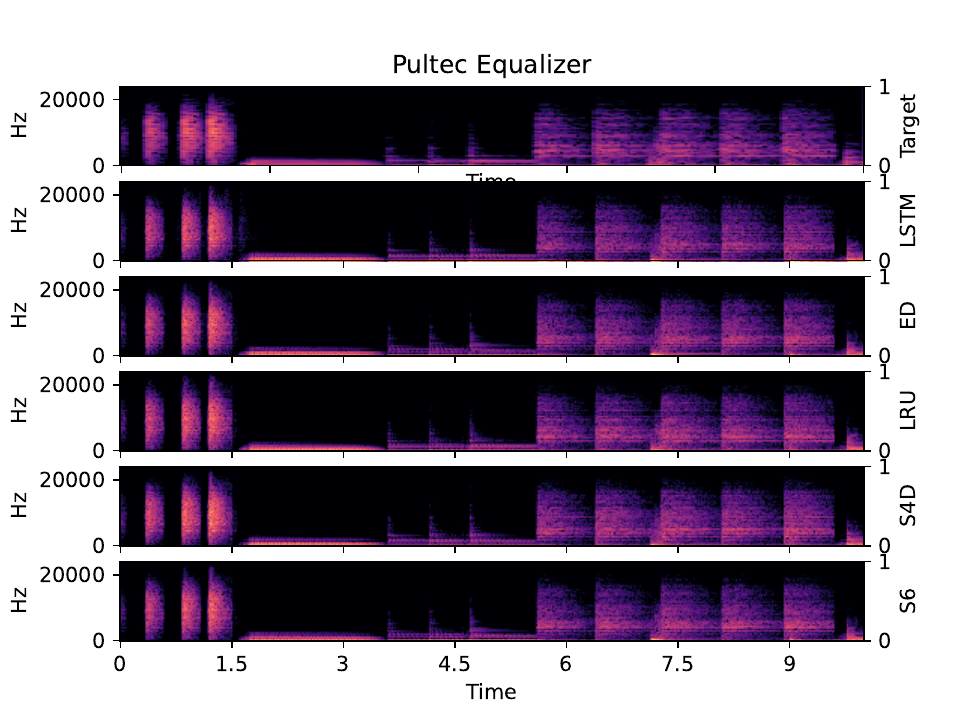}
    \includegraphics[scale=0.39]{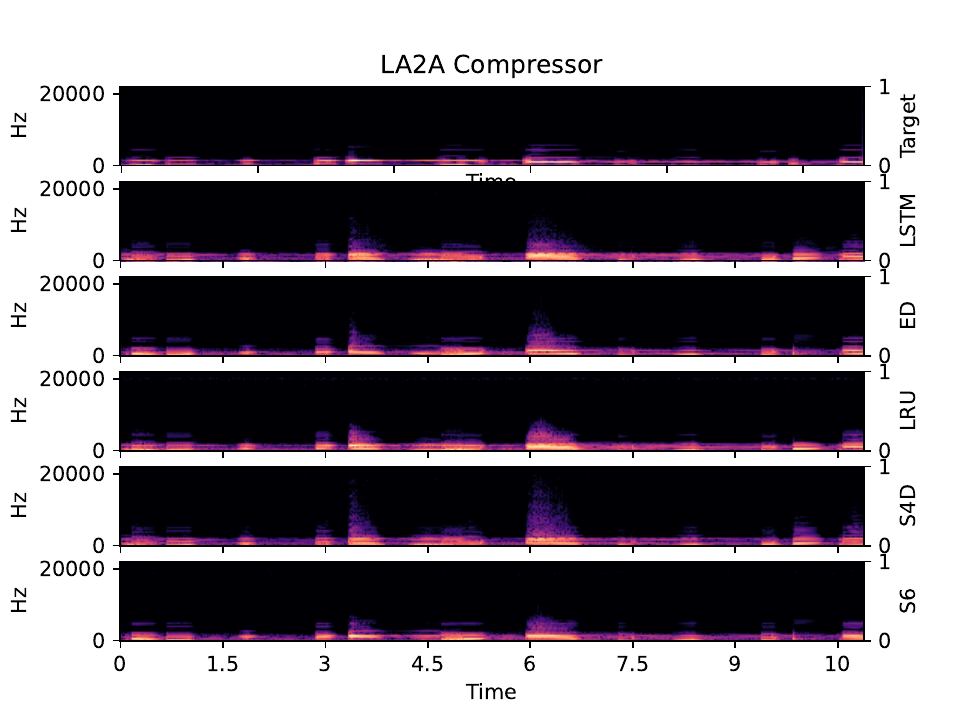}
    \includegraphics[scale=0.39]{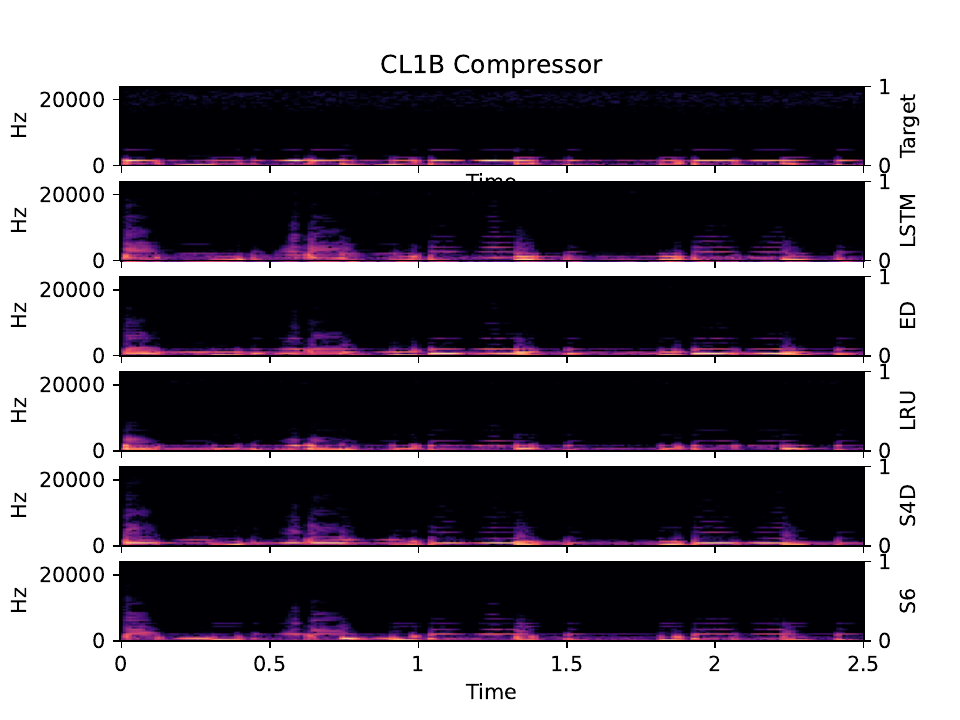}\\
    \includegraphics[scale=0.39]{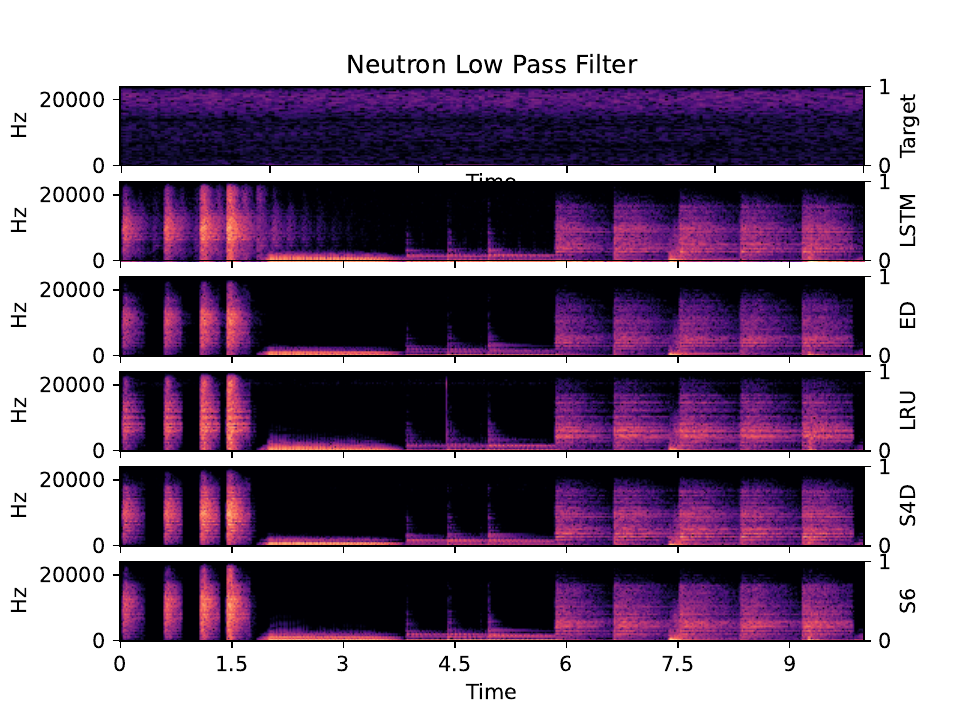}\\
    \end{center}
     \caption{Comparison of spectrograms of the target and predicted output. The STFT is computed using windows of $2048$ samples with $25\%$ overlap. The results refer to the highest conditioning values, representing the most challenging scenario.}\label{fig:stft}
\end{figure}

To further verify the model's capacity to learn the conditioned effect, we provide a series of plots that compare the target signal with the predictions generated by the five distinct architectures. For these plots, we chose the modeling scenario that we believe to be the most challenging based on the effect parameters used as conditioning factors for the networks. Except for the low-pass filter, this typically involves parameters at the maximum of the selected range since the output signal, under these conditions, is likely to deviate most significantly from the input. Regarding the low-pass filter, the parameters are configured for minimum cutoff and maximum resonance. The plots show the RMS energy in Figure ~\ref{fig:rmse}, and the spectrograms in Figure ~\ref{fig:stft}.
For datasets including conditioning parameters, the plots showing predictions versus the target are generated for a $10$ second input signal containing bass, guitar, and drum loop sound. These input signals are part of the respective test sets.

In the case of maximum distortion, we can notice \textbf{LRU} prediction having significantly lower energy than other models. Notably, the spectrograms shown in Figure~\ref{fig:stft} illustrate that \textbf{S4D} and \textbf{S6} models better match the energy in the harmonics at higher frequency regions. \textbf{LRU} fails to generate the harmonics in the OD300 case, while \textbf{ED} presents the opposite behavior in Neutron's overdrive. Models trained on the Neutron OD dataset exhibit larger errors, suggesting that this dataset presents a more challenging modeling case for the investigated architectures. This challenge can also be seen in the figures. The overdrive appears to be cutting the higher frequencies, behaving also as a low-pass filter. 
When inspecting models emulating the Saturator, \textbf{LRU} is again left behind, while other models present closer performances. In contrast, \textbf{LRU} achieves closer matching in the spectrograms, visible in Figure~\ref{fig:stft}. Other models tend to produce more harmonics, particularly the \textbf{LSTM}, which introduces noticeable distortions that are clearly audible in the audio output and adversely impact its performance. Models emulating the Equalizer demonstrate similar performance across different designs. However, when examining the spectrograms, \textbf{S6} shows superior matching to the target audio characteristics.

In the LA-2A case, we observe an overall good match for RMSE in Figure~\ref{fig:rmse}. The \textbf{LRU} model displays slightly more mismatches in both plots. Upon examining Figure ~\ref{fig:stft}, spectrograms related to the \textbf{S6} model have the best match with the target output, while the other models present similar mismatches when looking at the higher frequencies. More difficult is the CL 1B case, where \textbf{LRU} presents the best energy matching in Figure~\ref{fig:rmse}, and Figure ~\ref{fig:stft}. In the case of the compressor, the capability to encode past information of the signal is critical, and \textbf{LSTM} is confirmed to have a worse temporal tracking ability than the other models, which present design improvements. These improvements are particularly evident with the CL 1B when setting long attack and release times through the conditioning control parameters. These reach up to $300$ ms and $5$ seconds, respectively. 
In the case of the low-pass filter, input-output signal alterations are the most pronounced and diverse among the effects considered, challenging the network's ability to learn this mapping. The difficulty mainly stems from the filter's range, which varies from a straightforward pass-through at high cutoff frequencies to a highly selective filter at low cutoff frequencies. Nevertheless, informal listening of the predicted outputs indicates that all models struggle with accurately emulating the filter behavior; instead of eliminating high-frequency contents, they attenuate the whole signal, as apparent in Figure~\ref{fig:stft}. The \textbf{LSTM} model, in particular, tends to incorrectly predict excessive energy on the RMS energy, as visible in Figure~\ref{fig:rmse}, and to introduce distortions, as shown in Figure~\ref{fig:stft}.

Generally, performance metrics are relatively close across models across most effects examined in this study. When looking at objective metrics, no models appear clearly outperforming the others, but all share similar learning capability. MSE, ESR, and $M_{NRMSE}$ errors present the same trends, while performing better in the time domain does not lead to the same in the frequency domain. The \textbf{S4D} and \textbf{S6} models often yield similar performance, while \textbf{LRU} exhibits the greatest variance across different metrics and datasets. When modeling compressors, \textbf{S4D}, \textbf{S6}, and \textbf{ED} demonstrate high quality. The \textbf{ED}, \textbf{LRU}, and especially \textbf{S4D} and \textbf{S6} models excel at encoding historical signal information within their internal states, significantly outperforming \textbf{LSTM} models in compressor scenarios. Conversely, the \textbf{LSTM} models appear to perform well only when emulating effects with shorter and less critical transients and time dependencies, such as the overdrives. Furthermore, in certain instances, \textbf{LSTM} models tend to introduce significant distortion.

Table~\ref{tab:p} presents the $p$-values calculated using the Friedman test, which determines whether there are statistically significant differences between the medians of samples. All values are below $0.05$, indicating statistical significance in the results. The Wilcoxon results are omitted for brevity, as they involve computing $p$-values for each pair of models trained on the same dataset. However, we observe $p$-values exceeding $0.05$ solely when comparisons involve \textbf{LRU} or when models are trained with the low-pass filter dataset. This supports the previously mentioned difficulty in learning low-pass filters and the high variance in the \textbf{LRU} model's performance.

\begin{table}[h]
\caption{$p$-values calculated from the Friedman test for each metric and dataset presented in Table~\ref{tab:errors}, indicating the statistical significance of the rankings. }\label{tab:p}
\begin{tabular*}{\textwidth}{lccccc}
\toprule%
Dataset & MSE & ESR & $M_{NRMSE}$ & $M_{SF}$ & $M_{STFT}$\\
\midrule
OD300 & $0.001$ & $0.001$ & $0.0008$ & $0.011$ & $0.0008$\\
Neutron Overdrive & $0.001$ & $0.001$ & $0.0004$ & $0.0006$ & $0.0010$\\
Helper Saturator & $0.002$ & $0.002$ & $0.002$ & $0.001$ & $0.0004$\\
Pultec Passive EQ& $0.002$ & $0.002$ & $0.001$ & $0.0007$ & $0.013$\\
LA-2A & $0.002$ & $0.002$ & $0.012$ & $0.0006$ & $0.002$\\
CL 1B & $0.002$ & $0.002$ & $0.002$ & $0.001$ & $0.0004$\\
Neutron Filter & $0.002$ & $0.002$ & $0.004$ & $0.018$ & $0.002$\\

\end{tabular*}
\end{table}

\section{Conclusion}

Modeling analog effects is challenging, requiring simplifications to maintain numerical stability and efficiency. ANNs have gained popularity for their convenient, automated approach, but they can struggle with computational complexity and flexibility for control parameters. While hybrid solutions incorporating physics can offer advantages, black-box models are sometimes necessary. Recurrent neural networks, utilizing internal states for memory, provide interactive solutions by encoding signal history, enhancing their modeling capability beyond input size limitations. 

In this paper, we conducted a comparative analysis to assess the performance of LSTM models against recently proposed recurrent models, specifically LRU and SSMs, which have shown impressive results in sequence modeling. Five models were designed with parameter conditioning and low-latency response using the Feature-wise Linear Modulation method, focusing on effects like overdrive, saturation, low-pass filter, equalization, and compressors. Our evaluation employed metrics considering root-mean-squared energy, transients, and frequency content for valuable insights into future research.
Experimental results indicate that LSTM-based networks perform well in emulating distortion and equalizer effects. However, they tend to introduce noticeable distortions compared to other models in this comparative study. Conversely, LSTM structured in an encoder-decoder configuration, and SSMs, perform better in emulating tape saturation and compression effects, where encoding historical signal information may significantly impact the accuracy. However, the performance gap is not always pronounced. Regarding the low-pass filter, no architecture successfully learns its audio processing characteristics. In general, MSE, ESR, and $M_{NRMSE}$ errors present the same trends, while performing better in the time domain does not lead to the same in the frequency domain. Lastly, LRU shows inconsistent performance across different datasets.
\bibliographystyle{authordate1}
\bibliography{main}  

\end{document}